\documentclass[11pt,a4paper]{article}
\usepackage{jcappub}
\usepackage{slashed}

\newcommand{\csz}{c_{*0}}
\newcommand{\dsc}{\delta_\mathrm{sc}}
\newcommand{\Dsq}{\Delta^2}
\newcommand{\Dsqcb}{\Delta^2_\mathrm{CB}}
\newcommand{\Dsqone}{\Delta^2_\mathrm{1h}}
\newcommand{\Dsqtwo}{\Delta^2_\mathrm{2h}}
\newcommand{\Dvir}{D_\mathrm{v}}

\newcommand{\fnu}{f_\mathrm{N}}

\newcommand{\Heaviside}{\Theta}
\newcommand{\pst}{p_\mathrm{st}}
\newcommand{\qst}{q_\mathrm{st}}
\newcommand{\Reu}{R_\mathrm{E}}
\newcommand{\rhocb}{\bar\rho_\mathrm{CB}}

\newcommand{\rhovir}{\rho_\mathrm{v}}

\newcommand{\Rscale}{R_\mathrm{s}}

\newcommand{\Rvir}{R_\mathrm{v}}

\newcommand{\zta}{z_\mathrm{ta}}
\newcommand{\zvir}{z_\mathrm{v}}

\title{Spoon or slide?  The non-linear matter power spectrum in the presence of  massive neutrinos}
\author[a]{Steen Hannestad,}
\author[b]{Amol Upadhye,}
\author[b]{and Yvonne Y.\,Y.\,Wong}
\emailAdd{sth@phys.au.dk, a.upadhye@unsw.edu.au, yvonne.y.wong@unsw.edu.au}

\affiliation[a]{Department of Physics and Astronomy, Aarhus University\\
	Ny Munkegade 120, DK-8000 Aarhus C, Denmark}
\affiliation[b]{Sydney Consortium for Particle Physics and Cosmology\\
	 School of Physics, The University of New South Wales, Sydney NSW 2052, Australia}

\abstract{
  Numerical simulations of massive neutrino cosmologies consistently find a  spoon-like feature in the non-linear matter power spectrum ratios of cosmological models that differ only in the neutrino mass fraction $\fnu$.  Typically, the ratio approaches unity at low wave numbers~$k$, decreases by $\sim 10 \fnu$ at $k \sim 1\ h$/Mpc, and turns up  again at large $k$.    Using the halo model of large-scale structure, we show that this  spoon feature originates in the transition from the two-halo power spectrum to the one-halo power spectrum.  The former's sensitivity to $\fnu$ rises with $k$, while that of the latter  decreases with $k$. The presence of this spoon feature is robust with respect to different choices of the halo mass function and the halo density profile, and does not require any parameter tuning within the halo model.  We demonstrate that a standard halo model calculation is  already able to predict the depth, width, and position of this spoon as well as its evolution with redshift $z$ with remarkable accuracy.   Predictions at $z \gtrsim 1$ can be  further improved using non-linear perturbative inputs.

  }

\begin{document}

\maketitle

\section{Introduction}
\label{sec:introduction}
Following the 1998 discovery of atmospheric neutrino oscillations, the past two decades have seen an explosive proliferation of neutrino  experiments, detecting and measuring the properties of neutrinos from both natural and man-made sources~\cite{Tanabashi:2018oca}.  Of these, the totality of flavour oscillations data has enabled us to establish firmly (i)~maximal mixing amongst the three  standard-model families of neutrinos, and (ii)~that at least one neutrino state has a mass exceeding $\sim 0.05$~eV~\cite{deSalas:2017kay,Esteban:2018azc}.  Concurrently, laboratory kinematics constraints from weak decays --- notably the tritium $\beta$-decay endpoint measurements of the Mainz and Troisk experiments~\cite{Kraus:2004zw,Aseev:2011dq}  and most recently KATRIN~\cite{Aker:2019uuj}--- currently limit the effective electron neutrino mass to $m_{ee} \lesssim 1.1$~eV (90\%C.L.).  In combination, these experimental facts translate into a present-day cosmic neutrino energy density $\omega_\nu = \sum m_\nu/(94~{\rm eV})$  of $0.0005 \lesssim \omega_\nu \lesssim 0.04$, making  the  neutrino  an inevitable and potentially sizeable component of the dark matter.

Interestingly, cosmology itself also provides an independent constraint on $\omega_\nu$ and hence the neutrino mass sum $\sum m_\nu$ by way of 
the phenomenon of free-streaming and its associated impact on large-scale structure formation~\cite{Lesgourgues:2012uu,Wong:2011ip,Hannestad:2006zg}.    Within the framework of linear cosmological perturbation theory, the signatures of massive neutrino free-streaming in observables such as the cosmic microwave background (CMB) anisotropies and the large-scale matter power spectrum are well known and computed precisely by such Boltzmann solvers as~{\sc Camb}~\cite{Lewis:1999bs,Howlett:2012mh} and {\sc Class}~\cite{Lesgourgues:2011re}.
Indeed, an oft-repeated statement is that, comparing the present-day matter power spectrum of a massive to a massless neutrino cosmology, the former is 
suppressed on small scales by a fractional amount $8 \fnu$, where $\fnu := \omega_\nu/\omega_m$ is the fraction of the total matter density~$\omega_m$ in massive neutrinos.  Null observation of these  effects so far has allowed us to place an upper limit on $\sum m_\nu$ in the ball-park of $\lesssim 0.2$~eV for restrictive assumptions about the dark energy, though this weakens by a factor of $\approx 3$ when the dark energy equation of state and its derivative are simultaneously allowed to vary~\cite{Upadhye:2017hdl,Aghanim:2018eyx,RoyChoudhury:2019hls}.

Less precisely known, however, are the signatures of neutrino free-streaming in observables for which the dynamics of structure formation have become non-linear.  This is an especially pressing concern --- and one that has attracted growing interest and activity in recent years --- in view that
forthcoming cosmological  surveys such as the Large Synoptic Survey Telescope and the ESA Euclid mission are expected to derive most of their constraining power {\it vis-\'{a}-vis} neutrino masses ---  $1\sigma$ error forecasted at  $\sigma(\sum m_\nu) \approx 0.02$~eV~\cite{Hamann:2012fe,Amendola:2016saw} --- from observables of this category.
 Pioneering $N$-body simulations of massive neutrino cosmologies~\cite{Brandbyge:2008rv,Brandbyge:2008js,Brandbyge:2009ce} consistently found a maximum fractional power suppression that exceeds the linear-theory prediction of $8 \fnu$ --- approximately $10 f_N$ at $z=0$, attained at a wavenumber of  $k\sim 1\ h$/Mpc, a result that has been confirmed by subsequent, independent simulations~\cite{Viel:2010bn,Bird:2011rb,AliHaimoud:2012vj,Castorina:2015bma,Banerjee:2016zaa,Liu:2017now,Banerjee:2018bxy,Partmann:2020qzb}  and is likewise borne out by higher-order perturbative calculations~\cite{Wong:2008ws,Fuhrer:2014zka,Lesgourgues:2009am,Blas:2014hya,Upadhye:2015lia}.
	
Consensus, however, has yet to be reached on what should transpire beyond $k\sim 1\ h$/Mpc.  The $N$-body simulations of~\cite{Brandbyge:2008rv} observed a ``spoon'' feature in the massive-to-massless matter power spectrum ratio, wherein the neutrino-mass-induced fractional suppression first deepens to approximately $10 f_N$ at $k\sim 1\ h$/Mpc and then turns around, diminishing eventually to below even the linear-theory suppression of $8 \fnu$, as shown in figure~\ref{f:spoon}.  While this spoon feature has been repeatedly confirmed by independent N-body and hydrodynamical simulations~\cite{Bird:2011rb,AliHaimoud:2012vj,Banerjee:2016zaa,Pedersen:2019ieb,Partmann:2020qzb}, a confluence of factors surrounding the scale at which the spoon emerges has nonetheless cast lingering doubts in some quarters about its actuality:
%
%
%%%%%%%%%%%%%%%%
\begin{figure}[t]
	\begin{center}
		\includegraphics[width=150mm]{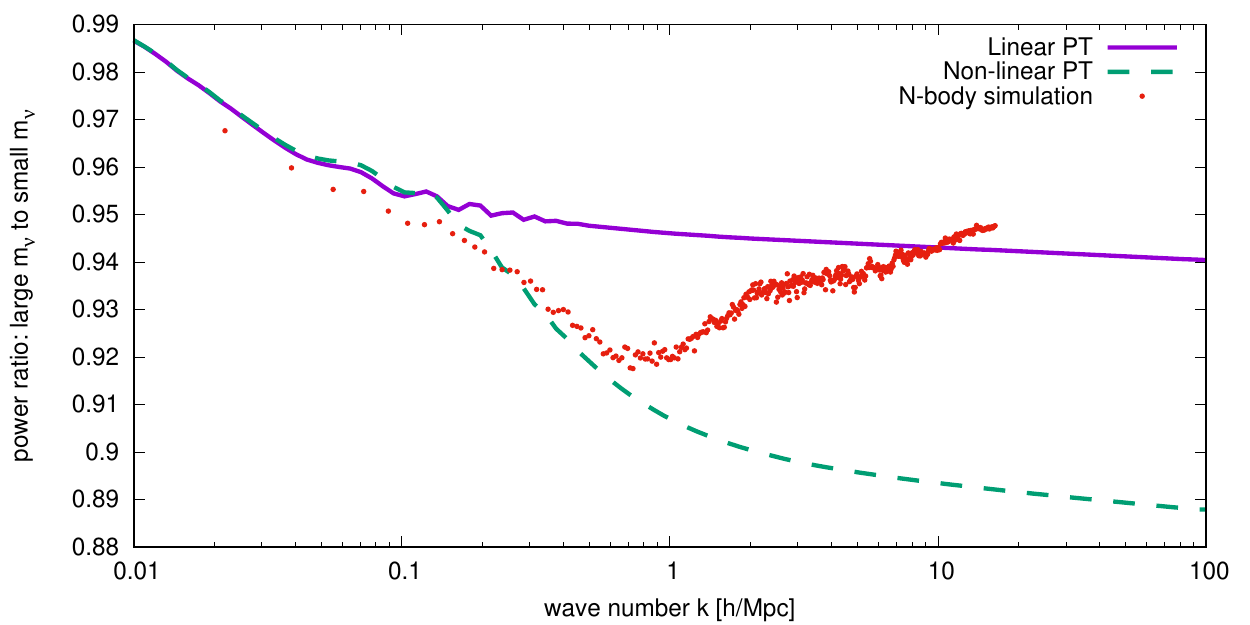}%
	\end{center}
	\caption{The spoon feature seen in $N$-body simulations is not evident
		in higher-order perturbative calculations of the matter power spectrum ratio,
		which prefer instead a ratio that continues to slide downwards at small scales.  The large-$m_\nu$ and small-$m_\nu$ models compared here correspond to $\fnu= 0.012$ and $\fnu=0.0044$ respectively.
		\label{f:spoon}
	}
\end{figure}
%%%%%%%%%%
%
%
\begin{enumerate}
\item No higher-order perturbative calculation to date has managed to reproduce the power spctrum spoon.  Rather, perturbative analyses prefer the fractional suppression to deepen further with increasing $k$, culminating in a ``slide''-shaped power spectrum ratio evident in figure~\ref{f:spoon}.

\item  Simulating the matter power spectrum at percent-level accuracy beyond $k \sim 1\ h$/Mpc is a computationally formidable task that necessitates the use of some of the largest computing facilities in the world.  While the spoon feature and  its upturn at $k \gtrsim 1\ h$/Mpc
pertain to the power spectrum {\it ratio}, for which achieving
percent-level numerical convergence is a much simpler affair~\cite{Hannestad:2019piu}, it remains a legitimate concern that the  large~$k$ upturn may be but an artefact of poor resolution and/or convergence.  This is especially so in view that simulations have not always produced spoons of the same shape or that dip at the same place.

\item  The manner in which massive neutrinos are represented in an 
$N$-body simulation may leave something to be desired on those same scales as well.  Shot noise is a known problem in those simulations that utilize a particle realization of the neutrino fluid, while hybrid methods that model neutrinos using some form of linear perturbation theory may be missing crucial non-linear physics.

\end{enumerate}

In this work, we investigate the spoon feature using an alternative approach based upon the halo model of large-scale structure~\cite{Cooray:2002dia}.  While other works such as reference~\cite{Massara:2014kba} have noted the presence of a spoon in halo models designed to fit realistic simulations, our goals are to establish the generality of the spoon within the halo model, and to quantify its depth, location, and variation with redshift and halo mass.  The premise of the halo model is that all clustering matter is contained within discrete units called ``halos''.  Then, describing the clustering statistics of the large-scale matter distribution reduces to stitching together several simple components: the matter distribution within a halo (halo density profile), the distribution of these halos in mass (halo mass function), and in space (halo bias). These components usually need to be established from and calibrated individually against simulations if precision is desired.  However, there exist also generic predictions following hierarchical structure formation arguments that are able to capture their qualitative behaviors to an acceptable level.

In connection to the matter power spectrum, the halo model description means that the two-point matter clustering statistics are on large scales dominated by correlations between two halos, and approach the linear-theory prediction in the region~$k \lesssim 0.1\ h$/Mpc, wherein massive neutrino suppression effects increase with $k$.   On small scales, the two-point statistics follow predominantly correlations within one halo.  Using generic (i.e., non-$N$-body calibrated) predictions for the halo mass function and density profile, we shall show that the one-halo correlation always has the opposite behavior relative to its two-halo counterpart:  in the one-halo term, a strong neutrino suppression prevails at low $k$ and then diminishes at high~$k$. The spoon shape observed in the total matter power  ratio originates thus in the transition from a {\it rising} two-halo power suppression to a {\it falling} one-halo one with increasing~$k$.  The dip in the spoon corresponds to a small-scale suppression of two-halo clustering, while the rise reflects the increasing independence of modes deeper inside a halo upon the background cosmology.

The rest of the paper is organized as follows.  Section~\ref{sec:background} introduces the main ingredients of the halo model. We compute in section~\ref{sec:one-halo_power_spectrum} the one-halo power spectrum using a variety of halo mass function and density profile inputs. These are combined with the two-halo power spectrum  and contrasted with predictions from $N$-body simulations in section~\ref{sec:halo_model_spoon}
to demonstrate the existence of the neutrino spoon.  We conclude in section~\ref{sec:conclusions}.

\section{Background}
\label{sec:background}
The halo model posits that all clustering matter in the universe is contained in halos.  It then follows that the simplest, two-point statistics of the large-scale matter distribution can be discussed in terms of correlations {\it within} a single halo and correlations {\it  between} two different halos.
In Fourier space, these two distinct contributions are respectively referred to as the (dimensionless) ``one-halo'' power spectrum $\Delta^2_{1 {\rm h}}(z,k)$ and ``two-halo'' power spectrum~$\Delta^2_{2 {\rm h}}(z,k)$, where
$\Delta^2(k) := \frac{k^3}{2\pi^2}P(k)$, and $P(k)$ is the Fourier transform of the two-point correlation function.
Together, these add to form the dimensionless total halo power spectrum
\begin{equation}
\Delta^2_{\rm halo}(z,k) = \Delta^2_{1 {\rm h}}(z,k) + \Delta^2_{2 {\rm h}}(z,k)
\end{equation}
of the clustering matter distribution  in the universe.
  We use the term ``clustering matter'' to mean cold dark matter (CDM) and baryons  --- collectively ``CB'' --- but not neutrinos, whose clustering around halos we neglect in this work.

The detailed forms of  $\Delta^2_{1 {\rm h}}(z,k)$  and $\Delta^2_{2 {\rm h}}(z,k)$ will be noted in the appropriate places.   Here, we discuss first in this section the main ingredients that make up these expressions.  See~\cite{Cooray:2002dia} for an authoritative review of the halo model formalism.

%%%%%%%%%%%%%%%%%%%%%%%%%%%%%%%%%%%%%%%%%%%%%%%%%%%%%%%%%%%%%%%%%%%%%%%%%%%%%%%%
\subsection{Spherical collapse of halos}
\label{subsec:spherical_collapse_of_halos}

Consider a spherical overdensity with comoving Lagrangian radius $R$ in an Einstein-de Sitter universe.  Let the linear-theory density contrast in the region be $\delta_0>0$ at the current time, so that at scale factor $a$ it is $a\delta_0>0$.  At early times $a \delta_0$ is much less than unity.  It then follows that the mass of the overdensity is well approximated by $M = \frac{4\pi}{3} \rhocb R^3$, where $\rhocb$ is the mean comoving density of clustering matter.

The comoving Eulerian radius $\Reu(z)$ of this region falls with time as described by the parametric equations
\begin{equation}
  \frac{\Reu}{R}
  =
  \frac{1+z}{\frac{5}{3}\delta_0} \, \frac{1-\cos\theta}{2};
  \qquad
  a(\theta) = \frac{1}{1+z} = \left(\frac{3}{4}\right)^{2/3}
  \frac{(\theta-\sin\theta)^{2/3}}{\frac{5}{3} \delta_0}.
  \label{eq:parametric}
\end{equation}
The physical radius $a\Reu$  therefore increases from $0$ at $\theta=0$ to its maximum value at $\theta=\pi$, turns around, and collapses back to zero at $\theta=2\pi$.    At turnaround, denoted by the turnaround redshift~$\zta := z(\pi)$, we find $(\Reu(\zta)/R)^3 = 16 / (9\pi^2)$.

Rather than collapsing all the way to zero, suppose that the overdensity virializes at a redshift~$\zvir := z(2\pi)$, with a physical radius $\Rvir$ equal to half the physical radius at turnaround, i.e., $a(2\pi) \Rvir = a(\pi) \Reu(\zta) / 2$.  The comoving size of this virialized object is then
\begin{equation}
  \left(\frac{\Rvir}{R}\right)^3
  = \left(\frac{1+\zvir}{2(1+\zta)} \frac{\Reu(\zta)}{R}\right)^3
  = \frac{1}{18\pi^2}
  =: \frac{1}{\Dvir},
  \label{eq:dvir}
\end{equation}
and has a constant comoving density equal to $3M/4\pi\Rvir^3 = \Dvir \rhocb \approx 178 \rhocb$, or, equivalently, a density contrast of $\Dvir - 1 \approx 177$.  We refer to such a virialized object as a halo.  The corresponding linear-theory density contrast at virialization --- the so-called linear collapse density contrast, $\delta_\mathrm{sc}$ --- is given by
\begin{equation}
  \dsc := a(2\pi) \delta_0 = \frac{3}{5} \left(\frac{3\pi}{2}\right)^{2/3}
  \approx 1.68647,
  \label{eq:dsc}
\end{equation}
where we have used equation~(\ref{eq:parametric}) to evaluate $a(2 \pi)$.

Note that extending the spherical collapse model to a $w$CDM or $\Lambda$CDM cosmology does in general yield $\dsc$ and $\Dvir$ values that differ from the Einstein-de Sitter predictions and are furthermore redshift-dependent.
  Importantly, however, within each class of cosmologies, adding a subdominant amount of scale-dependent growth to the spherical collapse treatment (due to, e.g., massive neutrinos or clustering dark energy) generally has no big impact on the outcome $\dsc$ and $\Dvir$: in the case of massive neutrinos, the fractional change is of order~$\fnu$~\cite{LoVerde:2014rxa}, while for clustering dark energy, the change is typically sub-percent~\cite{Basse:2010qp}.
 Henceforth, we shall use exclusively the Einstein-de Sitter values of $\dsc$ and $\Dvir$.

%%%%%%%%%%%%%%%%%%%%%%%%%%%%%%%%%%%%%%%%%%%%%%%%%%%%%%%%%%%%%%%%%%%%%%%%%%%%%%%%
\subsection{Halo density profiles}
\label{subsec:halo_density_profiles}

Suppose that a mass $M$ has virialized at comoving radius $\Rvir$ into a spherically symmetric halo with density $\rho(r)$ at a comoving distance of $r$ from its center.  Let $y := r/\Rvir$ and take for simplicity $\rho(y\Rvir) = \rhovir \varrho(y)$, where $\rhovir$ denotes the halo density at the virial radius.  Then,
\begin{equation}
M = 4\pi \rhovir \Rvir^3 \int_0^\infty {\rm d}y\,y^2 \varrho(y) =: 4\pi \rhovir \Rvir^3 \mu,
\end{equation}
where the dimensionless parameter $\mu$ depends on the shape of the assumed dimensionless density profile $\varrho(y)$ alone.

Finding the matter power spectrum requires that we calculate the normalized Fourier transform $U(\vec{k}) = M^{-1} \int {\rm d}^3r\,\exp(-i\vec k \cdot \vec r) \rho(r)$ of the halo profile.  For a spherically symmetric profile $\rho(r)$, the Fourier transform reduces to a Hankel transform and $U(\vec{k}) = U(k)$.
Furthermore, because of the simplified profile $\rho(y\Rvir) = \rhovir \varrho(y)$, the normalized Fourier transform~$U(k)$ is effectively dependent only on the dimensionless wave number $q := k \Rvir$, i.e., 
\begin{equation}
U(q) = \mu^{-1} q^{-1/2} \sqrt{\pi/2} \int_0^\infty {\rm d}y\,y^{3/2} J_{1/2}(qy) \varrho(y),
\end{equation}
 where $J_{1/2}$ denotes a Bessel function of the first kind.
Table~\ref{t:halo_density_profiles} shows $\varrho(y)$, $\mu$, and $U(q)$ for several profiles of interest.

%%%%%%%%%%%%%%%%
%%%%%%%%%%%%%%%%%
\begin{table}[t]
  \tabcolsep=0.2cm
  %\begin{footnotesize}
    \begin{center}
      \begin{tabular}{c|c|c|c}
\hline
\hline
        $\varrho(y)$
        &
        $\mu$
        &
        $U(q)$
        &
        Notes
        \\
        \hline
        \hline
        $\frac{\Heaviside(1-y)}{y^\alpha}$ ($\alpha<3$)
        &
        $(3-\alpha)^{-1}$
        &
        ${}_1F_2\left(\frac{3}{2}-\frac{\alpha}{2};\frac{3}{2},
        \frac{5}{2}-\frac{\alpha}{2}; -\frac{q^2}{4}\right)$
        & \\
        
        $\frac{\Heaviside(1-y)}{y^\alpha}$ ($\alpha=0$)
        &
        $1/3$
        &
        $\frac{3}{q^3}(\sin(q) - q\cos(q))$
        & tophat \\
        
        $\frac{\Heaviside(1-y)}{y^\alpha}$ ($\alpha=1$)
        &
        $1/2$
        &
        $\frac{2}{q^2} (1 - \cos(q))$
        &\\
        
        $\frac{\Heaviside(1-y)}{y^\alpha}$ ($\alpha=2$)
        &
        $1$
        &
        $\frac{1}{q} \mathrm{Si}(q)$
        & SIS\\
\hline
        $\exp(-cy)$
        &
        $2 c^{-3}$
        &
        $\frac{c^4}{(c^2+q^2)^2}$
        & \\

        $\exp(-c^2y^2/2)$
        &
        $c^{-3}\sqrt{\frac{\pi}{2}}$
        &
        $\exp\left(-\frac{q^2}{2c^2}\right)$
        & \\
\hline
        $\frac{\Heaviside(1-y)}{cy (1+cy)^2}$
        &
        $\log(1+c) - \frac{c}{1+c}$
        &
        $\frac{\sin(\frac{q}{c})
          \left[\mathrm{Si}(q+\frac{q}{c})-\mathrm{Si}(\frac{q}{c})\right]
          + \cos(\frac{q}{c})
          \left[\mathrm{Ci}(q+\frac{q}{c})-\mathrm{Ci}(\frac{q}{c})\right]
          - \frac{\sin(q)}{q+q/c}
          }
        {\log(1+c) - c/(1+c)}$
        & NFW \\
\hline
\hline
      \end{tabular}
    \end{center}
  %\end{footnotesize}
  \caption{Dimensionless halo density~$\varrho(y) = \rho(y\Rvir)/\rhovir$,
    mass~$\mu = M / (4\pi\rhovir\Rvir^3)$, and normalized Fourier-transformed
    density~$U(q)$ for several halo profiles.  Here, $y=r/\Rvir$ is the
    dimensionless radius, $q=k\Rvir$ the dimensionless wave number,
    $\Heaviside(x)$ the Heaviside step function, and $\mathrm{Si}(x)$
    and $\mathrm{Ci}(x)$ the sine and cosine integrals.
    \label{t:halo_density_profiles}
  }
\end{table}
%%%%%%%%%%%
%%%%%%%%%%%%

The first four rows of table~\ref{t:halo_density_profiles} list power-law density profiles truncated to zero at $r>\Rvir$.  Of particular interest here are (i)~the tophat halo ($\alpha=0$) which is a simple example of a profile with a smooth core, and (ii)~the Singular Isothermal Sphere (SIS; $\alpha=2$) which has a sharp central cusp.   In the fifth and sixth rows, the exponential and Gaussian profiles are not truncated at $\Rvir$, but have in each case an $r$-independent parameter $c$ that allows the density to fall on a length scale $\Rscale = \Rvir / c$ different from the virial radius.  A halo with large~$c$ has much of its mass concentrated inside a radius much smaller than $\Rvir$, for which reason $c$ is referred to as the concentration of the halo.  The seventh row displays the Navarro-Frenk-White (NFW) universal halo profile~\cite{Navarro:1996gj}, a reasonable fit to the dark matter halo profiles obtained from collisionless $N$-body simulations.  The concentration~$c$ of this last class of models generally correlates with the halo mass $M$ and redshift~$z$ with significant scatter~\cite{Bullock:1999he}.  However, parametrized fits exist for the mean concentration ${\bar c}(z,M)$, namely~\cite{Bullock:1999he},
\begin{equation}
{\bar c}(z,M) = \frac{\csz}{1+z} \left(\frac{M}{M_*(z)}\right)^{-0.13},
\label{e:concentration}
\end{equation}
where $\csz=9$, and $M_*(z)$ is a characteristic mass to be defined in section~\ref{subsec:halo_mass_function} under equation~(\ref{e:f_sheth-tormen}).
Unless stated otherwise, we shall use the fit~(\ref{e:concentration}) and neglect the scatter in $c(z,M)$ throughout this work.  Note however that 
 it is also possible to emulate ${\bar c}(z,M)$ from simulations across a broad range of cosmological parameters~\cite{Kwan:2012nd}.

%%%%%%%%%%%%%%%%%%%%%%%%%%%%%%%%%%%%%%%%%%%%%%%%%%%%%%%%%%%%%%%%%%%%%%%%%%%%%%%%
\subsection{Halo mass function}
\label{subsec:halo_mass_function}

The halo mass function quantifies the distribution of halos as a function of the halo mass~$M$.  Let $n(M)$ be the mean number density of halos with masses no greater than $M$.  Then, the dimensionful mass function $F(M) = {\rm d} n/{\rm d}M$ is the number density of halos per unit mass.  Since the halo model explicitly assumes that all clustering matter is contained in halos, the sum over all halo masses weighted by the halo mass function must equal the CDM+baryon energy density, i.e.,  $\int_0^\infty {\rm d}M\, M F(M) = \rhocb$.

While we generally expect the dimensionful halo mass function $F(M)$ to depend on cosmology, using the spherical collapse description of halos, it is possible to recast $F(M)$ into a ``universal'' form that is cosmology-independent at the $\sim 10$\% level.  To this end, we first define the mean-squared fluctuation amplitude inside spheres of comoving radius $R$ to be
\begin{equation}
  \sigma^2(z,R)
  =
  \int_0^\infty  {\rm d} \ln(k)\,  \Dsqcb(z,k) W^2(kR),
  \label{e:sigmaSq}
\end{equation}
where $\Dsqcb(z,k)$ denotes the dimensionless  linear CDM+baryon  power spectrum, and $W(x) = (3/x^3) \left[\sin(x) - x \cos(x) \right]$ corresponds to a real-space tophat filter.   Then, contrasting $\sigma(z,R)$ with the linear collapse density contrast $\dsc$ of equation~(\ref{eq:dsc}), we can generally expect the fraction of overdense regions collapsing into halos of mass $M = (4\pi/3) \rhocb R^3$ to be large if $\sigma(z,R) \gg \dsc$, while for  $\sigma(z,R)\ll \dsc$ halos of the corresponding masses will be rare.

At a given redshift $z$, let $\nu = \dsc / \sigma(R)$.  This implicitly defines $R(\nu)$ and hence $M(\nu)$
as monotonically increasing functions of $\nu$.%
\footnote{Note that the time-dependence of $R(\nu)$ characterizes changes in the population of halos with redshift.  It is not related to the dynamical evolution of the pre-virialization halo radius in the spherical collapse model discussed in section~\ref{subsec:spherical_collapse_of_halos}, which applies to a single halo.} 
Next, define 
\begin{equation}
f(\nu) = \frac{1}{\rhocb} M(\nu) F(M(\nu)) \frac{{\rm d}M}{{\rm d}\nu} ,
\end{equation} 
such that the requirement that all clustering matter be contained in halos translates to $\int_0^\infty f(\nu)\, {\rm d}\nu = 1$.  The function $f(\nu)$ is the so-called ``universal'' mass function, universal in the sense that  simulations have consistently shown $f(\nu)$ to be independent of cosmology --- including massive neutrino cosmologies --- and redshift at the $\sim 10\%$ level~\cite{Brandbyge:2010ge,Bhattacharya:2010wy,Biswas:2019uhy}.  In other words, to $\sim 10$\%-accuracy, cosmology and redshift affects only the mapping between $\nu$ and~$M$, but not the functional form of $f(\nu)$ itself.  Henceforth, we shall exclusively refer to $f(\nu)$ as the mass function.

Two commonly used mass functions are that of Press and Schechter~\cite{Press:1973iz} and that of Sheth and Tormen~\cite{Sheth:1999mn}, given respectively by
\begin{eqnarray}
  f(\nu)
  &=&
  \sqrt{\frac{2}{\pi}} \exp(-\nu^2/2),
 \qquad \hspace{44.7mm} \textrm{ (Press-Schechter),}
  \label{e:f_press-schechter}
  \\
  f(\nu)
 & =&
  A_\mathrm{st}(p_{\rm st},q_{\rm st}) \left(1 + \frac{1}{(\qst \nu^2)^{\pst}}\right) 
  \exp(-\qst\nu^2/2),
  \qquad \textrm{ (Sheth-Tormen).}
  \label{e:f_sheth-tormen}
\end{eqnarray}
The Press-Schechter mass function has no fitting parameters, while the more accurate Sheth-Tormen mass function has two, $\pst=0.3$ and $\qst=0.707$, obtained from fits to $\Lambda$CDM simulations, and a normalization $A_\mathrm{st}(p_{\rm st},q_{\rm st}) = 2^{\pst+1/2} \qst^{1/2} [2^{\pst}\pi^{1/2} + \Gamma(\frac{1}{2}-\pst)]^{-1} \approx 0.2162$.
Note that setting the Sheth-Tormen fitting parameters to $\pst=0$ and $\qst=1$ reproduces  the Press-Schechter mass function.
Both mass functions decrease exponentially for $\nu \gtrsim 1$, implying that halos of masses $M(\nu \gg 1)$ are exponentially rare.  It is useful to define a characteristic mass scale $M_* := M(1)$, to be interpreted as the largest mass at which halos are still common.  This (redshift-dependent) characteristic mass is the same $M_*(z)$ that appears in the parametrized fit~(\ref{e:concentration}) of the mean NFW halo concentration~\cite{Bullock:1999he}.

\section{Suppression of the one-halo power spectrum}
\label{sec:one-halo_power_spectrum}

The chief result of this work is that the power spectrum spoon arises through the transition from a two-halo power spectrum ratio whose dependence on $\fnu$ {\it rises} with $k$, to its one-halo counterpart whose $\fnu$-dependence {\it falls} with $k$.  Falling $\fnu$-dependence of the one-halo power spectrum implies that perturbations in the interior of each halo are less sensitive to the background cosmology than are perturbations on the outskirts of the halo.  The goal of this section is to demonstrate qualitatively the said behavior of the one-halo power spectrum ratio.  We begin with simple, analytical arguments in order to highlight the generality of our result, before proceeding to more accurate treatments incorporating well-motivated halo mass functions and  halo density profiles.  Throughout the article we plot CB power spectra and ratios, since observed galaxies trace the CDM and baryon fluids rather than the total matter.  Table~\ref{t:cosmological_models} lists the cosmological models used in this work.  

%%%%%%%%%%%%%%%%%%%%%%%%%%%%%%%%%%%%%%%%%%%%%%%%%%%%%%%%%%%%%%%%%%%%%%%%%%%%%%%%

\subsection{One-halo power spectrum}
\label{subsec:one-halo_integral}

\begin{figure}[t]
 	\begin{center}
 		\includegraphics[width=150mm]{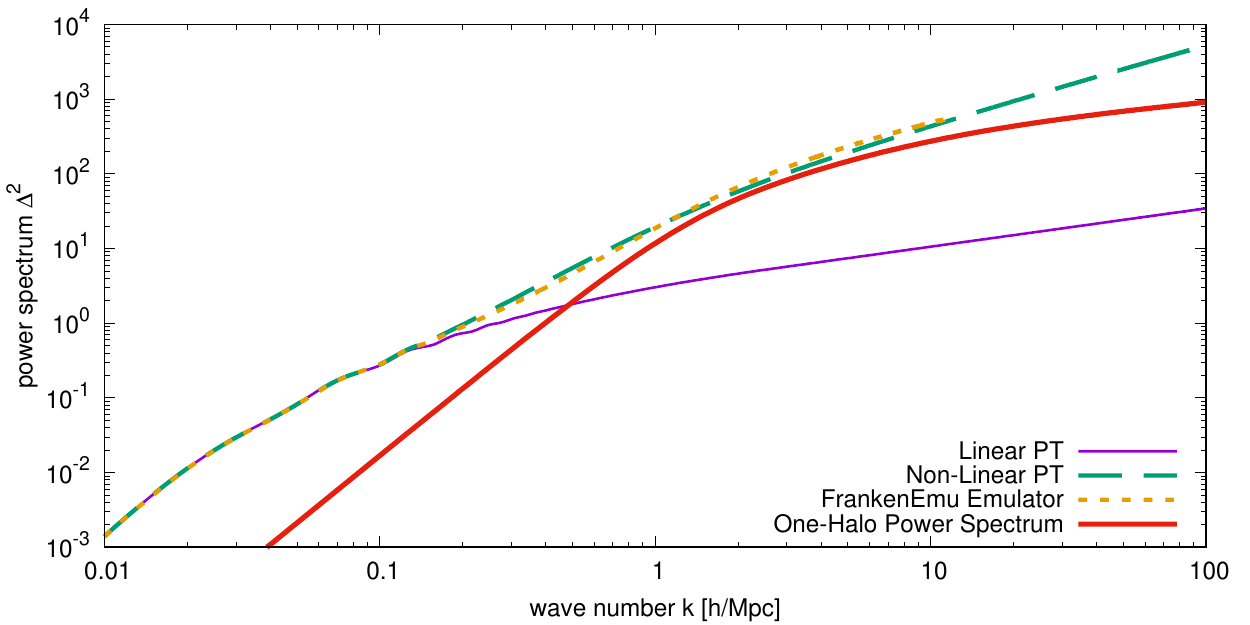}%
 	\end{center}
 	\caption{One-halo power spectrum for the $\Lambda$CDM(1) model of table~\ref{t:cosmological_models}, computed using the Sheth-Tormen mass function~(\ref{e:f_sheth-tormen}) and  the NFW halo density profile of  table~\ref{t:halo_density_profiles} and its associated mean concentration~(\ref{e:concentration}).  
 		 For comparison, we show also the corresponding linear power spectrum output of {\sc Camb}~\cite{Lewis:1999bs,Howlett:2012mh}, and  non-linear power spectrum predictions of Time-RG perturbation theory~\cite{Upadhye:2015lia,Upadhye:2017hdl} and of the FrankenEmu cosmic emulator~\cite{Heitmann:2013bra}.
 		\label{f:power_spectra}
 	}
 \end{figure}

The dimensionless one-halo power spectrum as defined in reference~\cite{Cooray:2002dia} can be written in our notation as
\begin{equation}
  \Dsqone(z,k)
  =
  \frac{2k^3}{3\pi} \int_0^\infty {\rm d}\nu f(\nu) R^3(z,\nu) U^2(z,k R(z,\nu) \Dvir^{1/3}).
  \label{e:Dsq1h}
\end{equation}
Here, the Fourier-space halo profile $U(z,q)$ of section~\ref{subsec:halo_density_profiles} has been recast as a function of $\nu$ as follows.  At a given redshift $z$, $\nu$ determines the comoving smoothing scale~$R(\nu)$ which we identify with the comoving Lagrangian radius $R$ of a spherical overdensity discussed in section~\ref{subsec:spherical_collapse_of_halos}.  It then follows from equation~(\ref{eq:dvir}) that 
 $\Rvir(\nu) = R(\nu)\Dvir^{1/3}$, and the dimensionless wave number $q$ on which $U$ depends is equivalently  $q = k R(\nu) \Dvir^{1/3}$.
As a point of reference, figure~\ref{f:power_spectra} shows the one-halo power spectrum for the $\Lambda$CDM(1) model of table~\ref{t:cosmological_models}, computed using the Sheth-Tormen mass function~(\ref{e:f_sheth-tormen}), the NFW density profile  of table~\ref{t:halo_density_profiles}, and the NFW mean concentration~(\ref{e:concentration}).

 %%%%%%%%%%%%%%%%%%
 %%%%%%%%%%%%%

Consider the limiting behaviors of $\Dsqone$.  At low $k$, the normalization of $U$ requires it to approach unity; the low-$k$ one-halo power spectrum  therefore  becomes
\begin{equation}
  \Dsqone(z,k\to 0) \to \frac{2 k^3}{3 \pi}\int  {\rm d} \nu \, f(\nu) R^3(z,\nu).
  \label{eq:onehalolowk}
\end{equation}
At the other end of the spectrum,  the high-$k$  behavior of $U$ depends upon our choice of halo profile.  Using for instance the truncated power law profiles of table~\ref{t:halo_density_profiles}, we find for the SIS profile $U(q \to \infty) \rightarrow \pi / (2q)$ and hence  
\begin{equation}
\Dsqone(z, k\to \infty) \rightarrow \frac{1}{6} \pi \Dvir^{2/3} k \int_0^\infty {\rm d} \, \nu f(\nu) R(z,\nu),
\label{eq:onehalohighk}
\end{equation}
 where,  as discussed in section~\ref{subsec:spherical_collapse_of_halos}, $\Dvir$ can be taken to be independent of $\nu$ and cosmology.  Thus, up to constant factors,  the one-halo integrand transitions from $fR^3$ at low $k$ to $fR$ at high~$k$ for the SIS halo.

%%%%%%%%%%%%%%%%%%%%%%%%%%%%%%%%%%%%%%%%%%%%%%%%%%%%%%%%%%%%%%%%%%%%%%%%%%%%%%%%
\begin{table}[t]
	\tabcolsep=0.2cm
	\begin{center}
		\begin{tabular}{r|ccccccc}
			\hline\hline
			&
			$\omega_m$
			&
			$\omega_b$
			&
			$\omega_\nu$
			&
			$h$
			&
			$A_s$
			&
			$n_s$
			&
			$\tau$
			\\
			
			\hline
			
			$\Lambda$CDM(1)
			&
			$0.1335$
			&
			$0.02258$
			&
			$0$
			&
			$0.71$
			&
			$2.1625\times 10^{-9}$
			&
			$0.963$
			&
			$0.09296$
			\\
			
			$\nu\Lambda$CDM(2)
			&
			$0.1335$
			&
			$0.02258$
			&
			$0.01$
			&
			$0.71$
			&
			$2.1625\times 10^{-9}$
			&
			$0.963$
			&
			$0.09296$
			\\
			
			$\nu\Lambda$CDM(3)
			&
			$0.1335$
			&
			$0.02258$
			&
			$0.001$
			&
			$0.71$
			&
			$2.1625\times 10^{-9}$
			&
			$0.963$
			&
			$0.09296$
			\\
			
			$\nu\Lambda$CDM(4)
			&
			$0.1432$%om
			&
			$0.0220$%ob
			&
			$0.000637$%on
			&
			$0.67$%h
			&
			$2.1\times 10^{-9}$%A_s
			&
			$0.96$%n_s
			&
			$0.09296$%tau
			\\
			
			$\nu\Lambda$CDM(5)
			&
			$0.1432$%om
			&
			$0.0220$%ob
			&
			$0.00171$%on
			&
			$0.67$%h
			&
			$2.1\times 10^{-9}$%A_s
			&
			$0.96$%n_s
			&
			$0.09296$%tau
			\\
			\hline
			\hline
		\end{tabular}
	\end{center}
	\caption{Cosmological models considered in this work,
		specified by the standard cosmological parameters: the total matter density $\omega_m := \Omega_m h^2$, which includes CDM, baryons, and neutrinos; the baryon density $\omega_b := \Omega_b h^2$; the neutrino density $\omega_\nu := \Omega_\nu h^2$; the dimensionless Hubble parameter $h = H_0 /$($100$~km/sec/Mpc); the amplitude $A_s$ of primordial scalar perturbations; the scalar spectral index $n_s$; and the optical depth $\tau$  to reionization.\label{t:cosmological_models}
	}
\end{table}

%%%%%%%%%%%

Indeed, this decrease in the exponent of $R$ in the integrand with increasing $k$  is a completely general feature of the one-halo power spectrum, arising from the generic expectation that $U$ must flatten at low $q$ and decrease at high $q$.  
If we were to approximate the high-$q$ behavior of $|U|$ as a power law,  i.e., $|U| \sim q^{-\beta}$, where $\beta$ is a positive constant, then the exponent of $R$ decreases from $3$ to $3-2\beta$.  Evidently from table~\ref{t:halo_density_profiles}, $\beta=1$ characterizes the high-$q$ behavior of the SIS halo, while $\beta=2$ describes the tophat.

Consider now a massive neutrino cosmology characterized by a neutrino fraction $f_N = \omega_\nu/\omega_m$.  If all other cosmological parameters besides $f_N$ are held fixed, then at high~$k$ we generically expect the linear CB power spectrum $\Dsq_{\rm CB}(z,k)$ to decrease with $f_N$ at  fixed $k$.   This decrease translates into a decrease in $\sigma^2(z,R)$ as defined in equation~(\ref{e:sigmaSq}) 
and hence an increase in $\nu$ at fixed~$R$.
Equivalently, because $R(\nu)$ is monotonically increasing, a reduction in linear power due to a finite $f_N$ results in a fixed $\nu$ mapping to a smaller value of $R$.

This mapping of $\nu$ to smaller $R$ values with increasing suppression in $\Dsq_{\rm CB}$ is precisely what we need to explain the $k$-dependence of the one-halo power spectrum ratio between a massive and a massless neutrino cosmology.
 Suppose that a neutrino-mass-induced suppression of the linear power spectrum maps to a fractional decrease of $\epsilon$ in $R$ at a fixed $\nu$.  Neglecting for now the $\nu$-dependence of $\epsilon$,  the one-halo power spectrum can be expected to reduce by $3\epsilon$ at low $k$ and by $(3-2\beta) \epsilon$ at high $k$, the latter of which evaluates to $\epsilon$ for the SIS halo.  Thus, through this simple estimate, we see that  suppression of the one-halo power spectrum due to massive neutrinos must decrease with increasing $k$.

%%%%%%%%%%%%%%%%%%%%%%%%%%%%%%%%%%%%%%%%%%%%%%%%%%%%%%%%%%%%%%%%%%%%%%%%%%%%%%%%
\subsection{An  analytical argument}
\label{subsec:a_simple_analytical_argument}

Before proceeding to a numerical integration of the one-halo power spectrum~(\ref{e:Dsq1h}) for different massive neutrino cosmologies, we first provide an analytical estimate of the expected fractional suppressions of the low-$k$ and high-$k$ one-halo powers in terms of the neutrino fraction~$f_{\rm N}$.
 For simplicity, we use in this subsection the Press-Schechter mass function~(\ref{e:f_press-schechter}) and the SIS halo profile of table~\ref{t:halo_density_profiles}.

The filter function~$W^2(kR)$ in equation~(\ref{e:sigmaSq}) acts as a sharp cutoff on the integrand at $k \approx 1/R$.  The effective integrand is thus sharply peaked, with a peak width given by the inverse of the effective spectral index $n_\Delta(k) := \partial \ln\Dsq_{\rm CB}/\partial \ln k$.  Thus, the integral~(\ref{e:sigmaSq}) evaluates approximately to 
\begin{equation}
\sigma^2(R) \approx \frac{\langle \Dsq_{\rm CB}(R^{-1})\rangle }{ \langle n_\Delta(R^{-1}) \rangle},
\label{eq:sigmapprox}
\end{equation} 
where the angle brackets $\langle \cdots \rangle$ serve as a reminder that integration normally smooths out the baryon acoustic oscillations present in $\Dsq_{\rm CB}$ and $n_\Delta$. In practice, to mimic this smoothing effect, we can simply replace $\Dsq_{\rm CB}$ and $n_\Delta$ with the no-wiggle power spectrum $\Dsq_{\rm eh}$ of Eisenstein and Hu~\cite{Eisenstein:1997ik} and its logarithmic derivative $n_{\Delta_{\rm eh}}$.  See details  in appendix~\ref{sec:eh}.

Armed with the approximation~(\ref{eq:sigmapprox}) and hence $\nu(R) \approx \dsc \left[n_{\Delta_{\rm eh}}(R^{-1}) / \Dsq_\mathrm{eh}(R^{-1})\right]^{1/2}$, 
we are now in a position to estimate the neutrino-mass-induced suppression to the low-$k$ and high-$k$ integrals as per equations~(\ref{eq:onehalolowk}) and (\ref{eq:onehalohighk}).  Suppose each integral is dominated by its integrand at $\nu=\nu_0$.  In order that $\nu_0$ remains the same under a change of cosmology, we must have 
\begin{equation}
\begin{aligned}
\frac{\delta \nu}{\nu_0}&  \approx \frac{1}{2} \left[\frac{\delta n_\Delta}{n_{\Delta,0}} - \frac{\delta \Dsq_{\rm CB}}{\Dsq_{\rm CB,0}} 
 \right] \\
 & = \frac{1}{2} \left\{ \epsilon_n(R_0^{-1}) - \epsilon_\Delta(R_0^{-1}) + \left[n_{\Delta,0} (R_0^{-1})-\left.\frac{\partial \ln n_{\Delta,0}}{\partial \ln k}\right|_{k=R_0^{-1}} \right] \frac{\delta R}{R_0}
  \right\} = 0,
  \label{eq:dnu}
 \end{aligned}
\end{equation}
where $\epsilon_n$ and $\epsilon_\Delta$ are fractional changes in  $n_\Delta$ and $\Dsq_{\rm CB}$, respectively, at fixed $k=R_0^{-1}$ due to neutrino masses.  
Typically, $|\epsilon_n| \ll |\epsilon_\Delta|$; the (smoothed) effective spectral index $n_\Delta$ also varies much more  slowly than the (smoothed) power spectrum $\Dsq_{\rm CB}$ with $k$.  Then, ignoring~$\epsilon_n$ and the derivative of $n_{\Delta,0}$,  equation~(\ref{eq:dnu}) can be rearranged to give
\begin{equation}
\frac{\delta R}{ R_0} \approx \frac{\epsilon_\Delta(R_0^{-1})}{n_{\Delta,0}(R_0^{-1})}
\end{equation}
for the fractional change in the mapping of $\nu$ to $R$ at $\nu= \nu_0$.

%%%%%%%%%%     
\begin{figure}[t]
	\begin{center}
		\includegraphics[width=150mm]{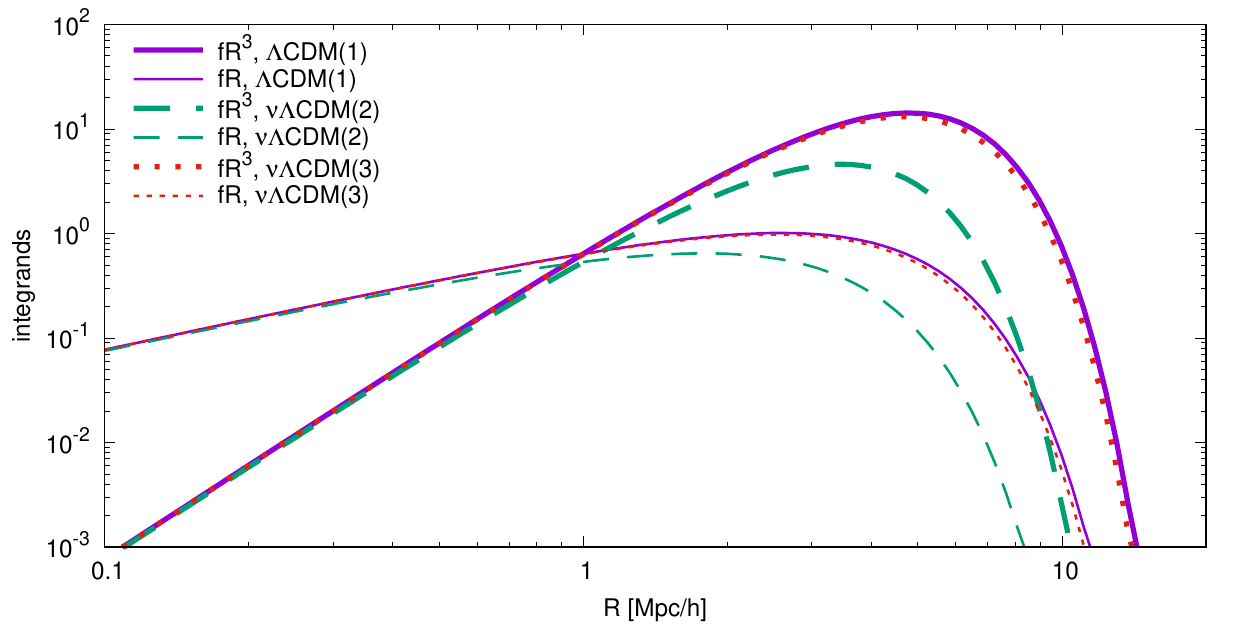}%
	\end{center}
	\caption{Integrands of the low-$k$ and high-$k$ integrals~(\ref{eq:onehalolowk}) and~(\ref{eq:onehalohighk}) as functions of $R$, evaluated in the approximation~(\ref{eq:sigmapprox}) using the Eisenstein and Hu no-wiggle power spectrum $\Dsq_{\rm eh}$ and the Press-Schechter mass function~(\ref{e:f_press-schechter}),  for three different cosmological models of  table~\ref{t:cosmological_models}.~\label{f:integrands_approx}
	}
\end{figure}
%%%%%%%%%%

Now, neutrinos making up a fraction $\fnu$ of the total matter changes the linear CB power spectrum $\Dsq_{\rm CB}$ by a fraction $\epsilon_\Delta \approx -6\fnu$  on scales relevant to the one-halo power spectrum~\cite{Lesgourgues:2012uu}.%
\footnote{While the {\it total} linear matter power spectrum --- defined as $\Dsq_m = (1-\fnu)^2 \Dsq_{\rm CB} + 2 (1-\fnu)\fnu \Dsq_{\rm CB \nu} + \fnu^2 \Dsq_{\nu}$, with $\Dsq_{\rm CB \nu}$ and $\Dsq_{\nu}$ the CB-neutrino cross and the neutrino power spectrum respectively  ---   is suppressed by a fraction $8 \fnu$ on small scales , the linear 
	CB power spectrum $\Dsq_{\rm CB}$ is suppressed only by $6 \fnu$.}
  It then follows that the low-$k$ integral~(\ref{eq:onehalolowk}) changes approximately by
\begin{equation}
\frac{\delta  \Dsqone(z,k\to 0)}{  \Dsq_{\rm 1h,0}(z,k\to 0)} \approx 3 \frac{\delta R}{ R_0} \approx - \frac{18\fnu}{n_{\Delta,0}(R_0^{-1})} \approx -12 \fnu, 
\label{eq:lowksupp}
\end{equation}
and the high-$k$ integral~(\ref{eq:onehalohighk}) by
\begin{equation}
\frac{\delta  \Dsqone(z,k\to \infty)}{  \Dsq_{\rm 1h,0}(z,k\to \infty)} \approx \frac{\delta R}{ R_0} \approx - \frac{6\fnu}{n_{\Delta,0}(R_0^{-1})} \approx -4 \fnu, 
\label{eq:highksupp}
\end{equation}
where in both cases we have used $n_{\Delta,0}(R_0^{-1}) \approx 1.5$ evaluated at the quasi-linear scale $k=0.1~h/$Mpc on the basis of $\Lambda$CDM(1) of table~\ref{t:cosmological_models}.   Actual numerical evaluations of the integrands for different cosmologies displayed in figure ~\ref{f:integrands_approx} reveal that they in fact peak at scales somewhat different from our canonical choice of $R_0^{-1} = 0.1~h$/Mpc.  Nonetheless, comparing the peak values of  $\Dsqone(z,k\to 0)$ and $\Dsqone(z,k\to \infty)$ between $\Lambda$CDM(1) and 
$\nu\Lambda$CDM(2) shows suppressions of $\approx 9 \fnu$ and $\approx 5 \fnu$ respectively; our estimates~(\ref{eq:lowksupp}) and (\ref{eq:highksupp}) are therefore reasonably good.

Thus, a rough picture of the non-linear power spectrum spoon emerges.  Going from small to large wave numbers~$k$, the two-halo power spectrum at first coincides with the linear CB power spectrum; here, the fractional power suppression due to massive neutrinos grows from zero at very low $k$ to $\approx 6 \fnu$ at larger, but still linear, wave numbers.  At some quasi-linear scale, the one-halo power spectrum begins to take over and eventually dominates; here, the fractional suppression first deepens to $\approx 9 \fnu$ and  then diminishes to $\approx 5\fnu$ at very large $k$ values.  This series of transitions,  $0 \to 6 \fnu \to 9 \fnu \to 5 \fnu$, in the fractional power suppression is the origin of the neutrino spoon.

%%%%%%%%%%%%%%%%%%%%%%%%%%%%%%%%%%%%%%%%%%%%%%%%%%%%%%%%%%%%%%%%%%%%%%%%%%%%%%%%
\subsection{Numerical computation}
\label{subsec:numerical_computation}

Having now developed an analytical picture of the neutrino spoon, we turn next to computing numerically the one-halo power spectrum~(\ref{e:Dsq1h}) in different massive neutrino cosmologies.  By default we use the Sheth-Tormen mass function~(\ref{e:f_sheth-tormen}), the NFW density profile  of table~\ref{t:halo_density_profiles}, and the NFW mean concentration~(\ref{e:concentration}) in our numerical computations.  Our goal in this subsection is to show that the decreasing $\fnu$-dependence of $\Dsqone$, and hence the spoon feature, is qualitatively stable over a broad range of halo profile and mass function choices.

%%%%%%%%%%%%%%%%%%%%%%%
\begin{figure}[t]
	\begin{center}
                \includegraphics[width=150mm]{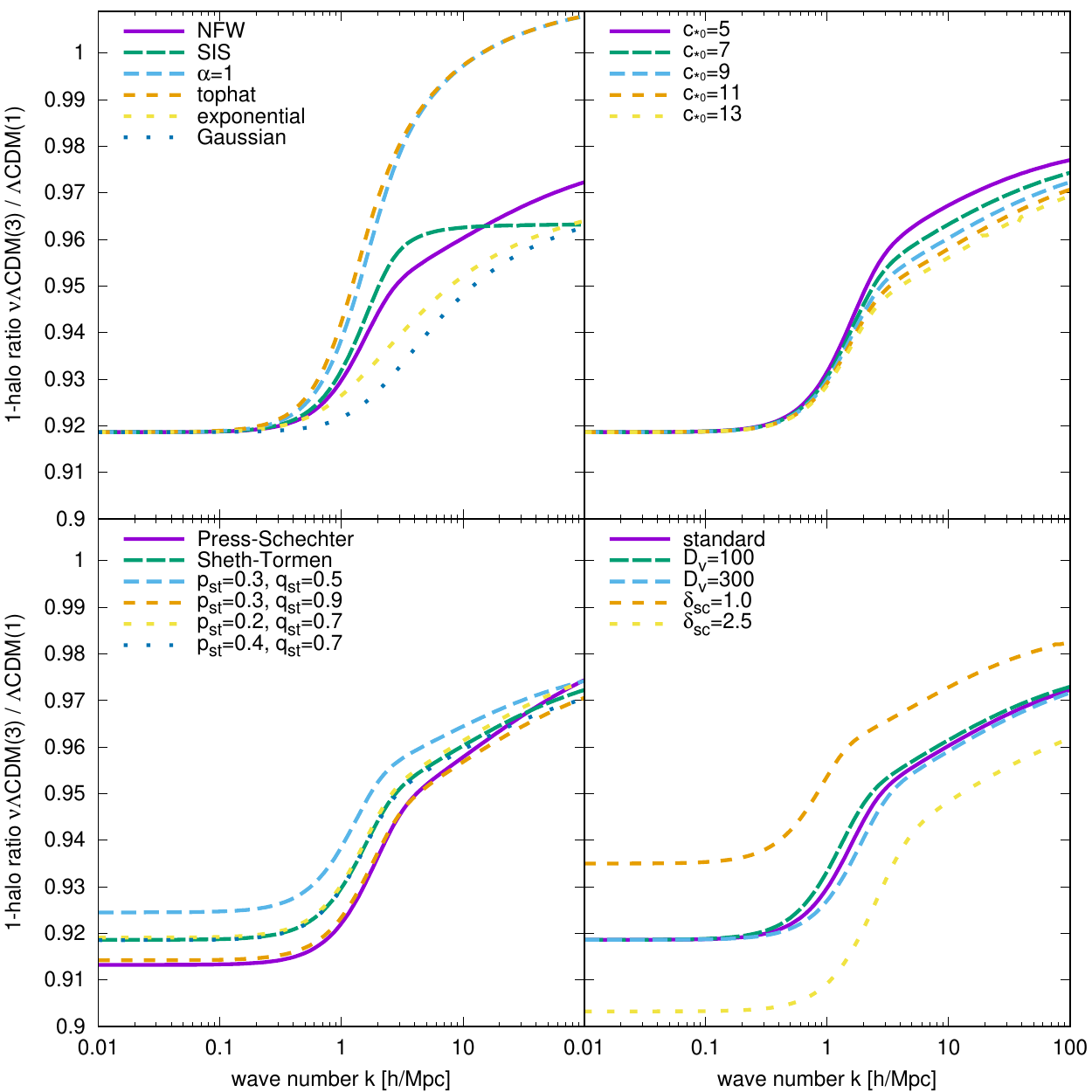}%
	\end{center}
	\caption{Ratios of the $\nu\Lambda$CDM(3) to the $\Lambda$CDM(1) one-halo power spectra, computed under variations of the halo model ingredients.  	Our default choices are the Sheth-Tormen mass function~(\ref{e:f_sheth-tormen}) and the NFW halo density profile of table~\ref{t:halo_density_profiles} and its associated mean concentration~(\ref{e:concentration}).   
{\it Top left}:~Variations in the halo profile following table~\ref{t:halo_density_profiles}.  {\it Top right}: Variations in the NFW concentration parameter~$c_{*0}$. {\it Bottom left}: Variations in 
the parameters	$\pst$ and $\qst$ of the halo  mass function~(\ref{e:f_sheth-tormen}).
{\it Bottom right}:~Variations in the virialization overdensity $\Dvir$ and the spherical collapse threshold
		$\dsc$.
		\label{f:ratio_1halo_profiles_varyUc}
	}
\end{figure}
%%%%%%%%%%%%%%%%%%%%%%%%%%

Figure~\ref{f:ratio_1halo_profiles_varyUc} shows the ratio of the $\nu\Lambda$CDM(3) to the $\Lambda$CDM(1) one-halo power spectrum under the following variations.
\begin{itemize}
\item {\it Top left}: We test all six halo density profiles  listed in table~\ref{t:halo_density_profiles};
\item {\it Top right}: We vary the NFW halo concentration parameter $c_{*0}$ in equation~(\ref{e:concentration}) over a range $c_{*0} \in [5,13]$;
\item {\it Bottom left}: We adjust the two parameters $\pst$ and $\qst$ of the Sheth-Tormen mass function~(\ref{e:f_sheth-tormen}) over the ranges $\pst \in [0,0.4]$ and $\qst \in [0.5,1]$, noting that the settings $\pst=0.3$ and $\qst=0.707$ correspond to the standard Sheth-Tormen
mass function, while $\pst=0$ and $\qst=1$ reproduces the Press-Schechter mass function; and

\item {\it Botton right}:  We tune the virialization overdensity $\Dvir$ and the spherical collapse threshold $\dsc$, one at a time, over the ranges  $\Dvir \in [100,300]$ and $\dsc \in [1.0,2.5]$.  
\end{itemize}

Consider first the top left plot of figure~\ref{f:ratio_1halo_profiles_varyUc}.  With $\fnu \approx 0.75\%$ for the $\nu\Lambda$CDM(3)  model, we see a large-scale suppression of $\approx 11 \fnu$ decreasing to a small-scale suppression of $\approx 5\fnu$ for the SIS profile; the NFW profile likewise yields a similar result up to  $k=10~h/$Mpc.  Thus, we predict an $\fnu$-dependent power spectrum suppression that increases from $0$ at large scales to~$6\fnu$ on quasi-linear scales, peaks at $11\fnu$ as the one-halo power spectrum comes to dominate, and then diminishes to $5\fnu$ at small scales.  This prediction is largely consistent with that of section~\ref{subsec:a_simple_analytical_argument} based on simple, analytical  estimates (section~\ref{subsec:a_simple_analytical_argument} finds a maximum suppression of~$9 \fnu$, as opposed to $11\fnu$ from numerical computation).

Secondly, figure~\ref{f:ratio_1halo_profiles_varyUc} reveals that  variations to the halo density profile (i.e., shape and $\Dvir$) affect primarily the high-$k$ behavior of spoon, while variations pertaining to the halo mass function (i.e., $\pst,\qst,$  and $\dsc$) impact on the power on all scales.  In all cases, however, the qualitative trend of a decreasing sensitivity to $\fnu$ with increasing $k$ is the same.  Importantly, the determination and calibration of {\it all}  ingredients used in the construction of the one-halo power spectrum --- the halo density profiles and  mass functions, the spherical collapse model, and even the halo model formalism itself --- predate the first observations of the neutrino spoon in $N$-body simulations~\cite{Brandbyge:2008rv,Brandbyge:2008js}. Thus, in this sense, one could argue that the neutrino spoon is but a natural consequence of hierarchical structure formation.

In summary, we have shown through analytical estimates and numerical computations that the sensitivity of the one-halo power spectrum to $\fnu$ falls with rising $k$.  This arises from the behavior of the Fourier-transformed halo profile $U(q)$, which is flat at low $q$ but decreasing at high $q$, and is robust with respect to our choice of mass function, density profile, and virialization parameters.  Physically, this decreasing sensitivity means that changes to the background cosmology have a greater effect upon fluctuations at the outskirts of a halo than upon fluctuations near its center.  This accords with our intuition that the virialization of a perturbation well inside a halo diminishes its memory of its initial condition, and makes it more dependent upon our choice of density profile, relative to a mode near the halo outskirts.  The top panels of figure~\ref{f:ratio_1halo_profiles_varyUc} provide further confirmation of this intuition.  For standard halo model parameter choices, we estimate the depth of the spoon as $\approx 10 \fnu$.  We will demonstrate in the next section that the decreasing $\fnu$-sensitivity of the one-halo term is essential for understanding the power spectrum spoon.

\section{Halo model spoon}
\label{sec:halo_model_spoon}

%%%%%%%%%%%%%%%%
\begin{figure}[tb]
	\begin{center}
		\includegraphics[width=150mm]{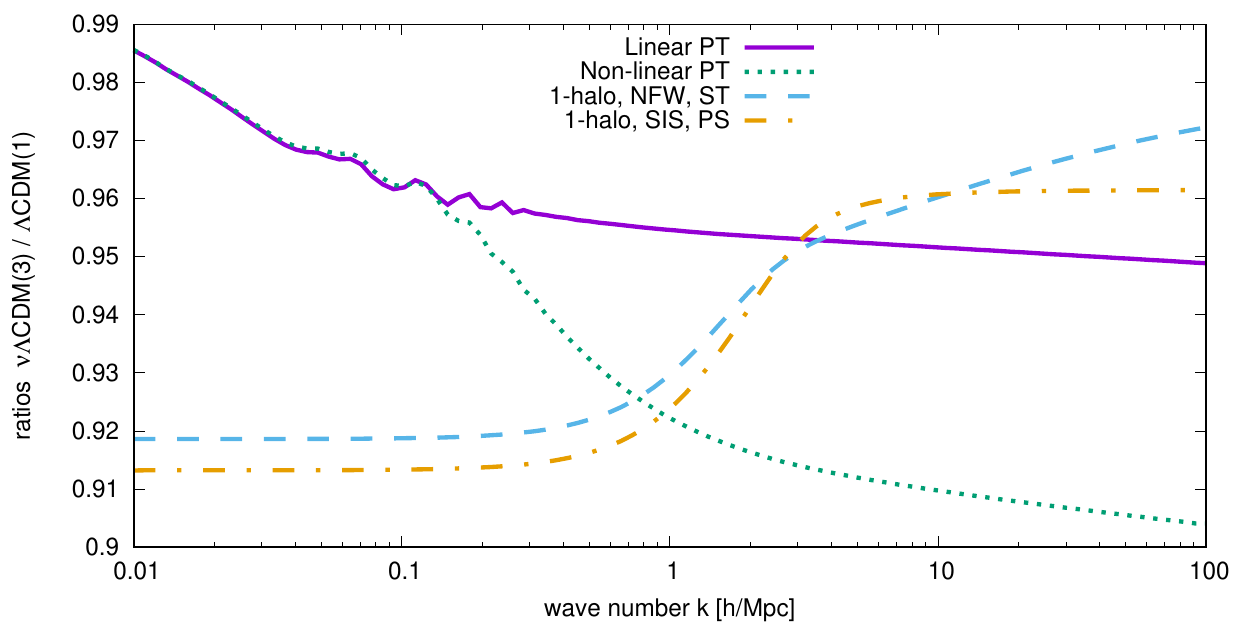}%
	\end{center}
	\caption{Ratios of the $\nu\Lambda$CDM(3) to the $\Lambda$CDM(1) CB power spectra, formed from (i) the linear power spectrum output of {\sc Camb}~\cite{Lewis:1999bs,Howlett:2012mh}, (ii) the non-linear predictions of Time-RG perturbation theory~\cite{Upadhye:2015lia,Upadhye:2017hdl}, and (iii) the one-halo power spectrum~(\ref{e:Dsq1h}) for two halo mass function and halo profile combinations.
		\label{f:ratios_1h_2h}
	}
\end{figure}
%%%%%%%%%%

Our central argument is that the neutrino spoon is born of (i) opposing behaviors of the two-halo and one-halo sensitivities to $\fnu$ as a function of $k$, and (ii) a transition from a two-halo to a one-halo dominance in the total CB power spectrum as we increase $k$.
Figure~\ref{f:ratios_1h_2h} motivates this argument.  Recall from figure~\ref{f:power_spectra} that the one-halo term becomes larger than the linear power at $k \approx 0.5~h/$Mpc.  Here, figure~\ref{f:ratios_1h_2h} shows that the neutrino sensitivity of the one-halo power spectrum drops below that of the linear power spectrum at $k\approx 4~h/$Mpc.  We therefore expect the bottom of the spoon to lie between these two wavenumbers.  In the following, we shall examine the location and magnitude of the spoon in more detail.

%%%%%%%%%%%%%%%%%%%%%%%%%%%%%%%%%%%%%%%%%%%%%%%%%%%%%%%%%%%%%%%%%%%%%%%%%%%%%%%%
\subsection{Two-halo power spectrum}
\label{subsec:two-halo_power_spectrum}

Following~\cite{Cooray:2002dia} the two-halo CB power spectrum can be approximated as
\begin{equation}
  \Dsqtwo(k) = B^2(k) \Dsqcb(k),
  \label{e:Dsq2h}
 \end{equation} 
where $\Dsqcb$ is the linear CDM+baryon power spectrum.
The scale-dependent bias factor~$B(k)$ is given by
\begin{equation}  
\begin{aligned}
  B(k)
  &=
  \int_0^\infty {\rm d}\nu\, f(\nu) U(z,k,M(z,\nu)) b_1(z,\nu),
  \label{e:Bias}
  \\
  b_1(z,\nu)
  &=
  1 +
  D(z) \cdot
  \left[\frac{\qst\nu^2-1}{\dsc}
    + \frac{2\pst}{\dsc \left(1 + (\qst\nu^2)^{\pst}\right)}\right],
    \end{aligned}
\end{equation}
where $D(z)$ is the linear growth factor, and $\pst$ and $\qst$ are the  fit parameters of the halo mass function~(\ref{e:f_sheth-tormen}).   On large scales, $B(k)$ approaches unity; on small scales, its effect is to turn off the two-halo power, since fluctuations at small separations likely belong in the same halo.

%%%%%%%%%%%%%%
\begin{figure}[tb]
	\begin{center}
		\includegraphics[width=150mm]{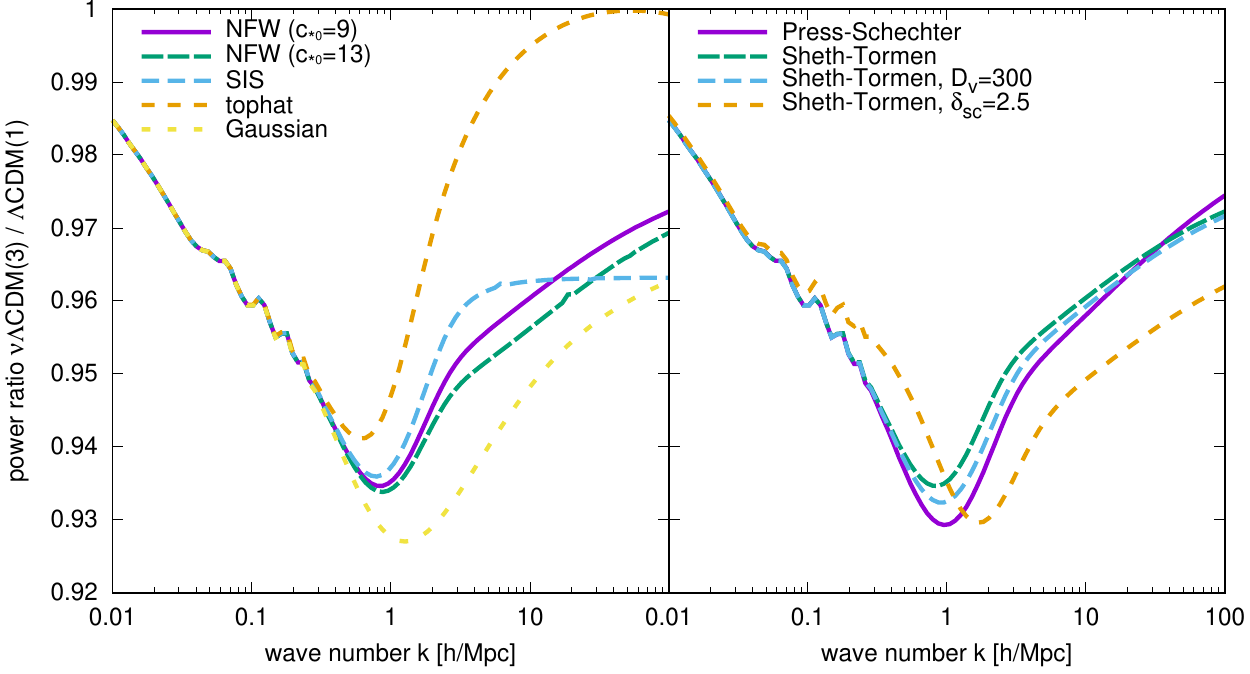}%
	\end{center}
	\caption{Ratios of the $\nu\Lambda$CDM(3) to the $\Lambda$CDM(1) total halo power spectra,  computed under variations of the halo model ingredients. 
		Our default choices are the Sheth-Tormen mass function~(\ref{e:f_sheth-tormen}), the NFW halo density profile of table~\ref{t:halo_density_profiles} and its associated mean concentration~(\ref{e:concentration}), and a two-halo term according to equation~(\ref{e:Dsq2h}),
		{\it Left}: Variations in the halo profile and the NFW concentration parameter $c_{*0}$.
		{\it Right}: Variations in the halo mass function parameters $\pst$ and $\qst$, virial density $\Dvir$, and spherical
		collapse threshold $\dsc$.
		\label{f:parameter-dependence}
	}
\end{figure}
%%%%%%%%%%%%%%

Combining equation~(\ref{e:Dsq2h}) and the one-halo term~(\ref{e:Dsq1h}), 
figure~\ref{f:parameter-dependence} shows ratios of the $\nu\Lambda$CDM(3) to the $\Lambda$CDM(1) total halo power spectra computed for a range of halo model variations.
   Consistent with figure~\ref{f:ratio_1halo_profiles_varyUc}, changing the halo profile and concentration affects primarily the small-scale behavior of the spoon, while  varying the mass function, $\Dvir$, and~$\dsc$ can broaden and deepen the spoon as well as shift the location of its minimum.   None of the tested variations, however, comes even close to eliminating the spoon feature altogether.  We therefore conclude that the existence of a spoon due to neutrino masses does not require any parameter tuning or special choices of halo model ingredients.

%%%%%%%%%%%%%%%%%%%%%%%%%%%%%%%%%%%%%%%%%%%%%%%%%%%%%%%%%%%%%%%%%%%%%%%%%%%%%%%%
\subsection{Redshift and halo mass dependence of the spoon}
\label{subsec:parameter-dependence_of_the_spoon}

\begin{figure}[tb]
	\begin{center}
          \includegraphics[width=150mm]{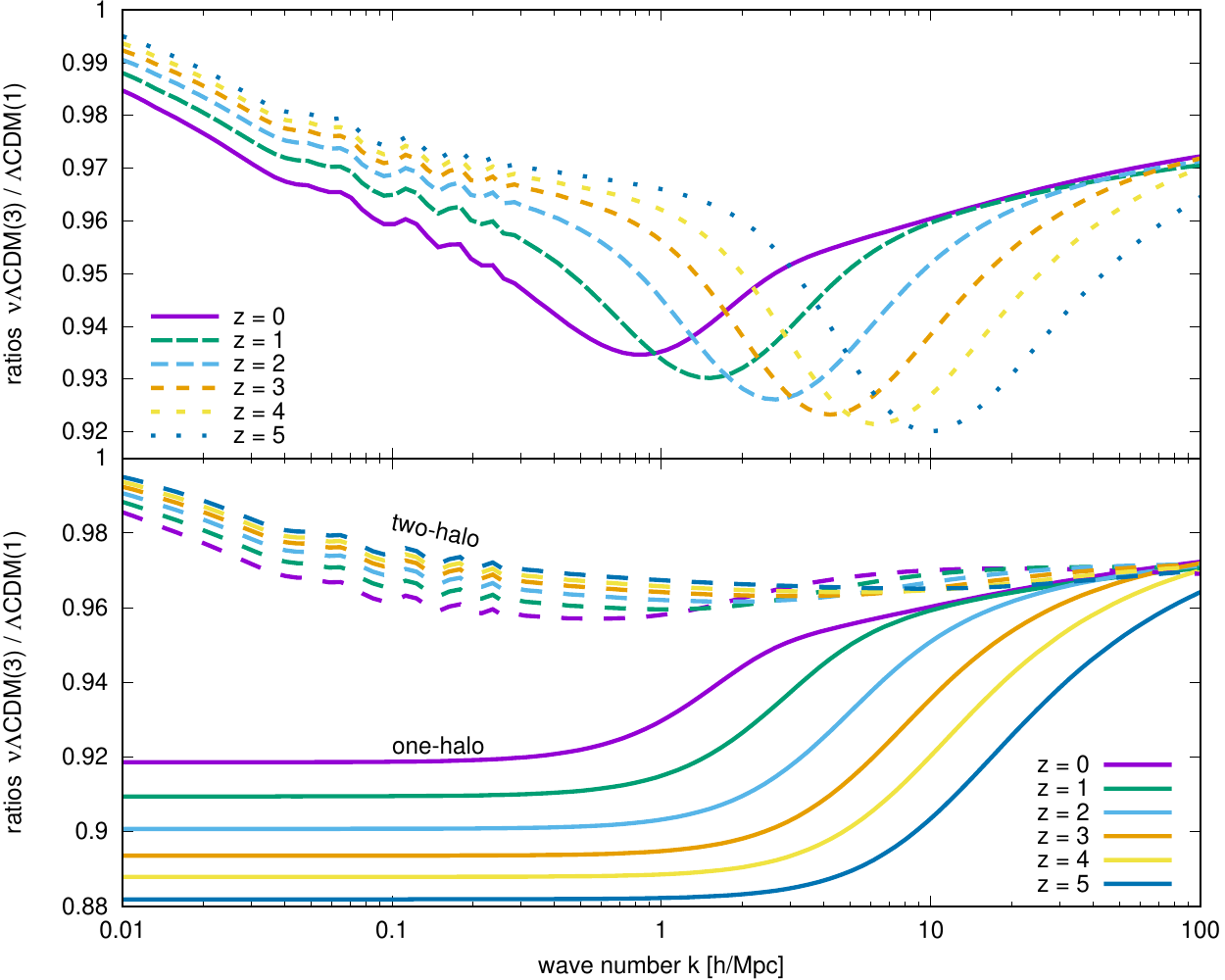}%
	\end{center}
	\caption{Ratios of the $\nu\Lambda$CDM(3) to the $\Lambda$CDM(1) power spectra at several redshifts, computed using the default Sheth-Tormen mass function, the NFW halo density profile, and a two-halo term according to equation~(\ref{e:Dsq2h}).   {\it Top}: Ratios of the total power spectra according to equation~(\ref{e:Dsq2h}).  {\it Bottom}: Ratios of one-halo (solid) and two-halo (dashed) power spectra.
		\label{f:spoon_vary_z}
	}
\end{figure}

Let us also consider the redshift dependence of the spoon in figure~\ref{f:spoon_vary_z}.  Two trends are evident in the top panel.  Firstly, the spoon shifts to smaller scales at higher redshifts.  This comes about because  the linear growth factor $D(z)$ decreases at higher $z$, causing $R(\nu)$ at fixed $\nu$ to decrease, thereby pushing the one-halo power to smaller scales.  This rightward shift of the one-halo power is particularly apparent in the bottom panel of  figure~\ref{f:spoon_vary_z}, and  is consistent with our expectation that, at higher redshifts, non-linear corrections to the power spectrum come to prominence on smaller scales.

%%%%%%%%%%%%%%%%%%
%%%%%%%%%%%

Secondly, the spoon deepens at higher redshifts.  Recall from section~\ref{subsec:a_simple_analytical_argument} that reducing~$R$ at which the one-halo integrand is maximized also forces us to evaluate the effective spectral index $n_\Delta(k)$ in equations~(\ref{eq:lowksupp}) and (\ref{eq:highksupp}) at a higher value of $k= R^{-1}$. Because $n_\Delta(k)$ decreases with $k$ and is fairly independent of redshift, this immediately means that the neutrino-induced one-halo power suppressions (\ref{eq:lowksupp}) and (\ref{eq:highksupp}) must evaluate to larger magnitudes
as seen in the bottom panel of figure~\ref{f:spoon_vary_z},
which in turn deepen the spoon.   Thus, our simple analytical approximation of section~\ref{subsec:a_simple_analytical_argument} is qualitatively consistent with both trends in figure~\ref{f:spoon_vary_z}.

%%%%%%%%%%%%%%%%%%
\begin{figure}[tb]
	\begin{center}
          \includegraphics[width=150mm]{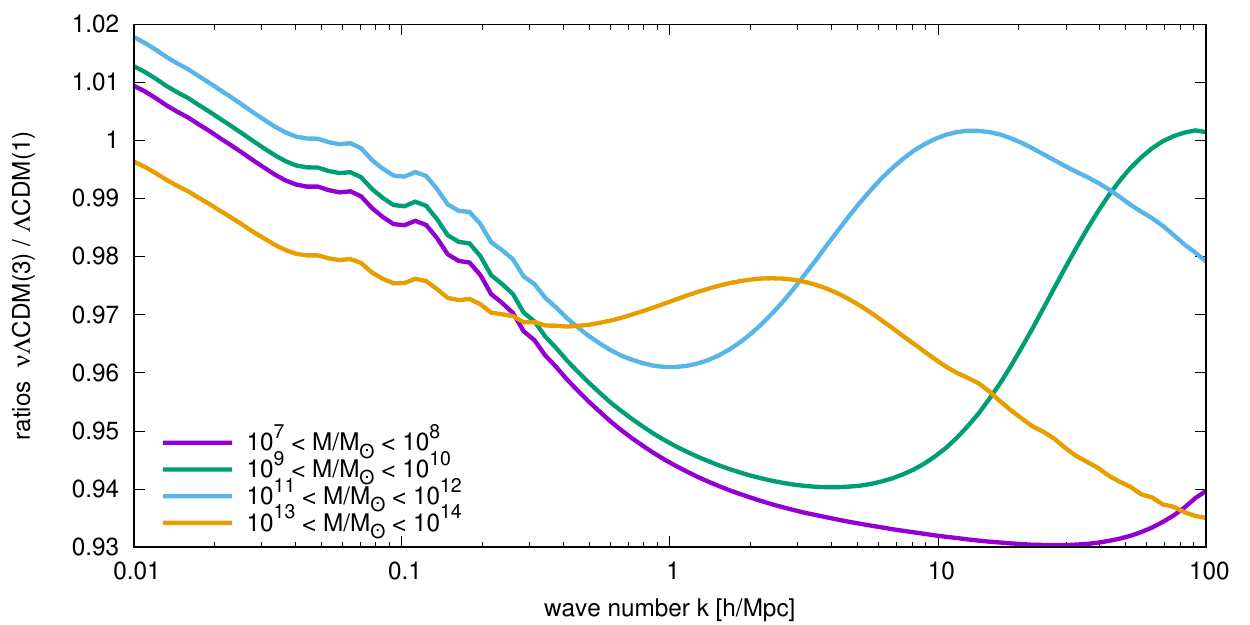}%
	\end{center}
	\caption{Ratios of the $\nu\Lambda$CDM(3) to the $\Lambda$CDM(1) power spectra for several halo mass ranges, computed using the default Sheth-Tormen mass function, the NFW halo density profile, and a two-halo term according to equation~(\ref{e:Dsq2h}), at redshift $z=0$.
	  \label{f:spoon_vary_M}
	}
\end{figure}
%%%%%%%%%%%

Next, we consider the dependence of the spoon on the halo mass.  Thus far, we have integrated over all $\nu$ --- hence all halo masses --- in the one-halo power spectrum~(\ref{e:Dsq1h}) and the two-halo bias~(\ref{e:Bias}).  Figure~\ref{f:spoon_vary_M} restricts these integrals to several narrow ranges of halo masses.  We find that each mass range has its own spoon, with lower-halo-mass spoons being deeper and shifted to smaller scales.  Since different types of galaxies trace different halo mass ranges at small scales, an observational search for a spoon must take care to identify the halo masses being probed.  Additionally, the mass-dependence of the spoon is a novel prediction of our calculation, which may be tested by observing multiple galaxy populations.

%%%%%%%%%%%%%%%%%%%%%%%%%%%%%%%%%%%%%%%%%%%%%%%%%%%%%%%%%%%%%%%%%%%%%%%%%%%%%%%%
\subsection{Comparison with $N$-body simulations}
\label{subsec:n-body_simulations}

Finally, we test the precision of our halo model spoon predictions against $N$-body simulations.  Massive neutrinos introduce new systematic effects into simulations, which are presently under better control  for small neutrino masses ($\sum m_\nu \lesssim 0.5$~eV) than for larger ones~\cite{Brandbyge:2008js}.
In order to minimize the impact of these systematic effects on our calculation, we choose to compare two low-mass neutrino models, $\nu\Lambda$CDM(4) and $\nu\Lambda$CDM(5) from table~\ref{t:cosmological_models}.  Model $\nu\Lambda$CDM(4) has $\sum m_\nu = 0.059$~eV, near the lower bound implied by neutrino oscillations in a normal neutrino hierarchy~\cite{Esteban:2018azc}, while $\nu\Lambda$CDM(5) has $\sum m_\nu = 0.159$~eV, at the upper end of the range currently allowed by cosmological measurements in the simplest extensions to the base $\Lambda$CDM model~\cite{Aghanim:2018eyx}. The remaining parameters of $\nu\Lambda$CDM(4) and $\nu\Lambda$CDM(5) are either identical or close to the best-fit values inferred from the Planck 2018 data~\cite{Aghanim:2018eyx}.

Our set of high resolution $N$-body simulations has been performed using a modified version of the {\sc Pkdgrav3}~\cite{Potter:2016ttn} code including general relativistic effects and a grid-based implementation of massive neutrinos as described  in Dakin {\em et al.}~\cite{Dakin:2017idt}. First discussed in~\cite{Brandbyge:2008js},  the grid-based implementation 
 incorporates only linear neutrino perturbations, and can thus 
compute the power spectra to 1\% precision or better strictly
only for those cosmologies with modest neutrino masses, $\sum m_\nu \lesssim 0.5$~eV~\cite{Brandbyge:2008js}, a criterion satisfied by both $\nu\Lambda$CDM(4) and $\nu\Lambda$CDM(5).  The simulations are performed in a box of side length $L = 384$~Mpc$/h$ with $N=1024^3$ cold particles.  These settings enable us to extract a power spectrum spanning the wave numbers $0.05 \lesssim k/(h/{\rm Mpc}) \lesssim 15$.  All runs have been initialized at $z = 99$ using the output of {\sc Class}~\cite{Blas:2014hya} in the $N$-body gauge.
 
%%%%%%%%%%%%%
\begin{figure}[t]
  \begin{center}
    \includegraphics[width=150mm]{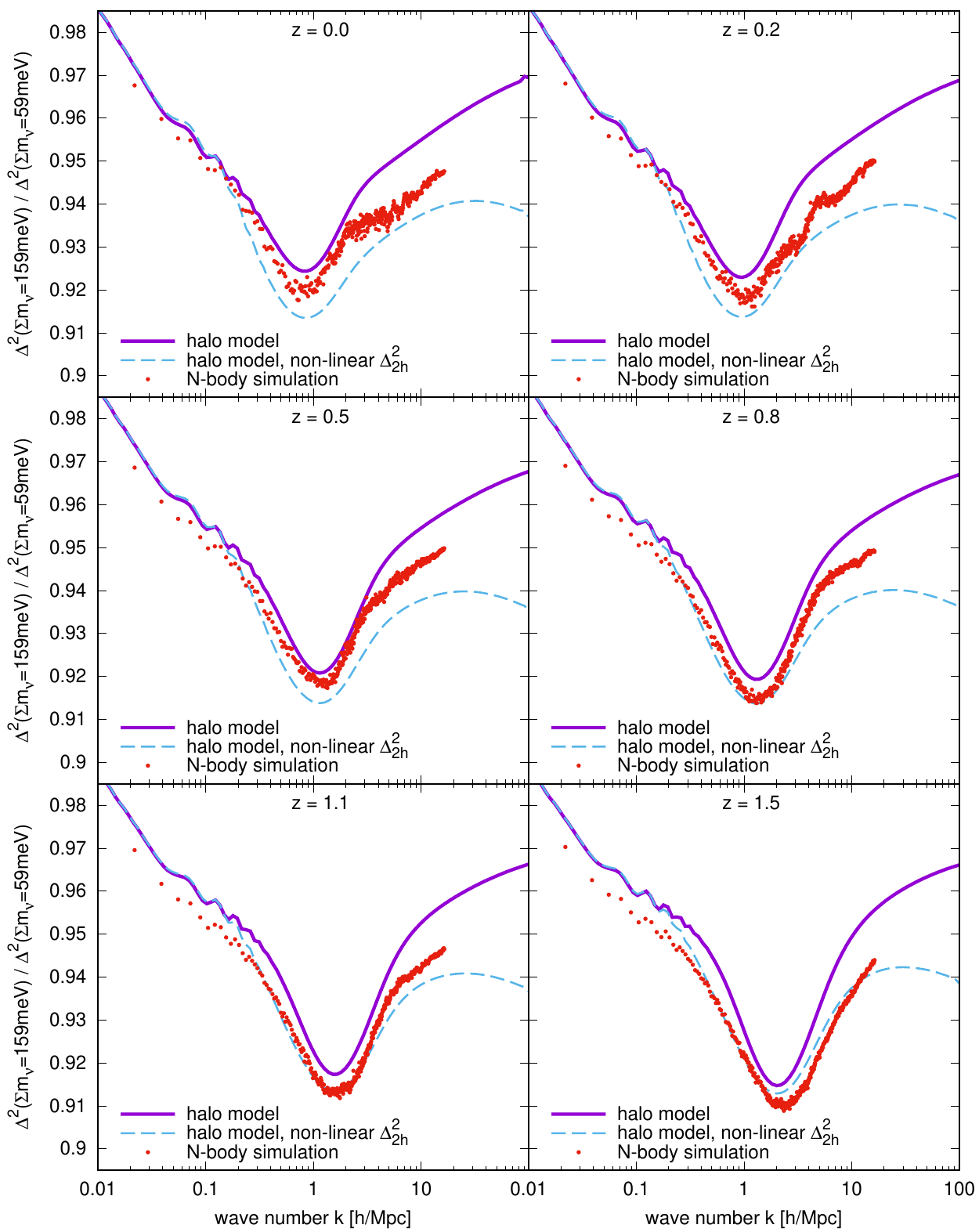}%
  \end{center}
\caption{Ratios of the $\nu\Lambda$CDM(5) to the $\nu\Lambda$CDM(4) power spectra at several redshifts, obtained from (i)~$N$-body simulations, (ii)~our default halo model  using the default Sheth-Tormen mass function, the NFW halo density profile, and a two-halo term according to equation~(\ref{e:Dsq2h}), and (iii)~same as~(ii), but with the linear-theory $\Dsqcb$ in the two-halo term~(\ref{e:Dsq2h}) replaced with its counterpart computed from Time-RG perturbation theory.    \label{f:spoons_halo_vs_Nbody}
  }
\end{figure}
%%%%%%%%%%%%%5

Figure~\ref{f:spoons_halo_vs_Nbody} shows the ratios of the $\nu\Lambda$CDM(5) to the  $\nu\Lambda$CDM(4) CDM+baryon power spectra at redshifts $z=0$ to $z=1.5$ obtained from our simulations.  Compared with our default halo model calculations --- which, we emphasise again, use the standard spherical collapse parameters $\Dvir=18\pi^2 \approx 178$ and $\dsc= (3\pi/2)^{2/3}\cdot 3/5 \approx 1.68647$, the NFW halo profile, and the Sheth-Tormen mass function, without further adjustments --- we see good agreement between the two sets of predictions in terms of both the position and depth of the spoon, as well as its deepening and rightward shift with increasing redshift.  
 
 Interestingly, reference~\cite{Cooray:2002dia} suggests that using a non-linear perturbative CDM+baryon power spectrum rather than the linear one as input in the two-halo term~(\ref{e:Dsq2h}) may improve the halo model predictions.  We consider this option in the context of the neutrino spoon, and replace $\Dsqcb$ in the two-halo term~(\ref{e:Dsq2h}) with its counterpart computed from the Time-RG perturbation theory of references~\cite{Pietroni:2008jx,Lesgourgues:2009am} as implemented in references~\cite{Upadhye:2015lia,Upadhye:2017hdl}.

 As shown in figure~\ref{f:spoons_halo_vs_Nbody}, using a non-linear perturbative $\Dsqcb$ input improves the agreement between the halo model and $N$-body simulations in terms of predicting the broadening and deepening of the spoon at higher redshifts ($z \gtrsim 1$); it however overestimates both of these effects at lower redshifts.  This finding is qualitatively consistent with those of references~\cite{Upadhye:2013ndm,Upadhye:2015lia} that  up to $k \sim 1\ h$/Mpc,  Time-RG perturbation theory yields results compatible at the $10\%$ level with $N$-body simulations at $z \gtrsim 1$ but overpredicts non-linear growth at $z \lesssim 0.5$.  At these lower redshifts, our linear and non-linear perturbative calculations provide approximate upper and lower bounds, respectively, on the $N$-body spoon.  Further accuracy may be achievable through non-linear corrections to the halo bias, as in references~\cite{McDonald:2006mx,Nishizawa:2012db,Desjacques:2016bnm}.  However, since these corrections involve divergent integrals whose regularization introduces fitting parameters into the power spectrum, we do not pursue this approach here.

 We therefore conclude that a simple halo model calculation accurately predicts the existence of the $N$-body neutrino spoon, as well as its depth, width, location, and redshift evolution.  Discrepancies between the two at $z \gtrsim 1$ can be addressed through non-linear perturbation theory with no further adjustment to the halo mass function and density profile.  Remaining low-redshift differences are consistent with the breakdown of perturbation theory seen in other comparisons with simulations.

%%%%%%%%%%%%%%%%%%%%%%%%%%%%%%%%%%%%%%%%%%%%%%%%%%%%%%%%%%%%%%%%%%%%%%%%%%%%%%%%
\subsection{Comparison with the literature}
\label{subsec:comparison_with_the_literature}

The presence of the power spectrum spoon has been observed in simulations over the years~\cite{Brandbyge:2008rv,Bird:2011rb,AliHaimoud:2012vj,Massara:2014kba,Banerjee:2016zaa,Pedersen:2019ieb,Partmann:2020qzb}.  Notably, Massara, {\em et al.}~\cite{Massara:2014kba} attributes the spoon to a greater reduction in the number of large halos relative to small ones in massive neutrino cosmologies.  Although our argument in section~\ref{subsec:a_simple_analytical_argument} is qualitatively consistent with their finding that a reduction in power spectrum amplitude gives rise to a spoon, we disagree with their explanation in two important details.

Firstly, the two-halo power spectrum term depends upon two powers of the mass function while the one-halo term depends upon just one.  If the spoon results from a greater reduction in the number of large halos relative to small halos with increasing $\fnu$, then the two-halo term ought to be the dominant contributor to the total spoon.  Evidently from the lower panel of figure~\ref{f:spoon_vary_z}, such a two-halo spoon is indeed present.  However, it is several times smaller than the actual spoon observed in $N$-body simulations and becomes shallower at higher redshifts,  both of which effects  are incompatible with the two-halo term being the dominant source of the spoon feature.  By contrast, in section~\ref{sec:one-halo_power_spectrum} we explain the spoon using the decreasing $\fnu$-dependence of the one-halo power spectrum with increasing $k$, which implies that the background cosmology has a greater effect on fluctuations farther from the halo center.   Our one-halo explanation of the spoon correctly estimates its depth, about $10 \fnu$, and its deepening with redshift in figure~\ref{f:spoon_vary_z}.

Secondly, if the spoon feature did arise through the greater dependence of the power spectrum on low-mass halos at small scales, then we would not expect such a feature in the power spectrum of halos of a single mass or of masses within a narrow range.  By contrast, as shown in figure~\ref{f:spoon_vary_M}, 
we do predict such spoon features within several narrow ranges of halo masses, 
 provided that (i)~the halos trace the linear power spectrum on large scales, and that (ii)~perturbations deep inside each halo are less sensitive to the cosmological background than perturbations on its outskirts.   This is an observationally testable prediction, since different types of galaxies are found in halos of different masses.

\section{Conclusions}
\label{sec:conclusions}
We have demonstrated that the spoon feature seen in the power spectrum ratio of massive-to-massless neutrino (or high-$m_\nu$ to low-$m_\nu$) cosmologies is a generic prediction of the halo model of large-scale structure, which assigns all clustering, cold matter to virialized structures called halos.  Through analytical and graphical estimates as well as numerical calculations, we have shown that the power spectrum associated with a pair of density fluctuations {\it within one halo} has a decreasing sensitivity to the neutrino mass fraction $\fnu$ with increasing wave number~$k$.  This contrasts with the power spectrum for a pair of density fluctuations belonging to {\it two different halos}, whose $\fnu$-dependence increases with increasing~$k$.  Thus, beginning at the largest length scales and considering successively smaller scales, we see a power spectrum ratio initially equal to unity falling with $k$ due to neutrino suppression of two-halo clustering, and then rising again as virialization causes perturbations deep inside each halo to ``forget'' their initial conditions.

This qualitative trend, the origin of the power spectrum spoon, is remarkably robust with respect to variations of the halo mass function, the halo density profile, and the spherical collapse model.  As demonstrated in figures~\ref{f:ratio_1halo_profiles_varyUc} and \ref{f:parameter-dependence},  notwithstanding wide variations of the halo model parameters well beyond standard values, we were unable to find a parameter region in which the one-halo power exhibits the undesired behavior of becoming more $\fnu$-dependent on small scales.

Our simple halo model prediction, based upon the Sheth-Tormen mass function and the Navarro-Frenk-White halo profile, matches $N$-body simulations remarkably well in figure~\ref{f:spoons_halo_vs_Nbody}.  Moreover, at redshifts $z \gtrsim 1$, residual differences between our predictions and simulations can be substantially reduced by using non-linear perturbation theory as input for the computation of the two-halo power spectrum.  Though we do not expect even non-linear perturbation theory to predict accurately the absolute matter power spectrum  at $k \gtrsim 1~h/$Mpc, evidently the associated errors are sufficiently $\fnu$-independent that they largely cancel out of the power spectrum ratio, yielding accurate calculations of the power spectrum spoon that do not require any parameter tuning whatsoever.%

In conclusion, the $N$-body power spectrum spoon in massive neutrino cosmologies observed repeatedly over the past twelve years is a real, physical phenomenon rather than a systematic error associated with the simulations themselves.  It arises through a non-perturbative effect, namely, the collapse and virialization of cold matter into  halos.  The spoon feature is confirmed to remarkable accuracy by a simple halo model calculation (with optional higher-order perturbative inputs), which reproduces $N$-body predictions of the depth, width, position, and redshift-dependence of the spoon with no need for fitting parameters.

\acknowledgments

AU and  Y$^3$W are  supported by the Australian Research Council's Discovery Project (project DP170102382) and Future Fellowship (project FT180100031) funding schemes.  The authors are grateful to J.~Dakin and J.~Kwan for insightful conversations, and thank J.~Dakin for assistance with the {\sc Pkdgrav} software.

\appendix

\section{Eisenstein and Hu no-wiggle power spectrum}
\label{sec:eh}
The Eisenstein and Hu no-wiggle linear power spectrum and the corresponding effective spectral index are given by~\cite{Eisenstein:1997ik}
 \begin{eqnarray}
\Dsq_\mathrm{eh}(z,k)
&=&
\frac{\mathcal N}{2\pi^2} k^{3+n_s} D^2(z) T^2_\mathrm{eh}(k),
\label{e:Dsqeh}
\\
n_{\Delta_{\rm eh}}(z,k) &:=& \frac{\partial \ln \Dsq_{\rm eh}}{\partial \ln k} =
3 + n_s + 2\, \frac{\partial \ln T_\mathrm{eh}(k)}{\partial \ln k}.
\label{e:n_Delta}
\end{eqnarray} 
Here, $D(z)$ is the linear growth factor (equal to the scale factor~$a$ in an Einstein-de Sitter universe), $\mathcal N$ is a normalization constant with units $(\mathrm{Mpc}/h)^{3+n_s}$, and $T_\mathrm{eh}(k)$ is the linear transfer function.  The last is well fitted by
  \begin{equation}  
  T_\mathrm{eh}(k)
      =
  \frac{L_\mathrm{eh}(k)}{L_\mathrm{eh}(k) + C_\mathrm{eh}(k) q_\mathrm{eh}^2(k)},
  \end{equation}
 with coefficients
\begin{equation}
\begin{aligned}
  L_\mathrm{eh}(k)
  &=
  \ln(2e + 1.8 q_\mathrm{eh}(k)),
  \\
  C_\mathrm{eh}(k)
  &=
  14.2 + \frac{731}{1 + 62.5 q_\mathrm{eh}(k)},
  \\
  q_\mathrm{eh}(k)
  &=
  \frac{k}{\Gamma_\mathrm{eff}(k)}
  \left(\frac{T_\mathrm{CMB,0}}{2.7~\mathrm{K}}\right)^2,
  \\
  \Gamma_\mathrm{eff}(k)
  &=
  \frac{\omega_m}{h}
  \left(\alpha_\Gamma + \frac{1-\alpha_\Gamma}{1 + (0.43 k s)^4}\right),
  \\
  \alpha_\Gamma
  &=
  1 - 0.328\frac{\omega_b}{\omega_m}\ln(431\omega_m)
  + 0.38 \left(\frac{\omega_b}{\omega_m}\right)^2 \ln(22.3\omega_m),
  \\
  s
  &=
  \frac{44.5 h \ln(9.83/\omega_m)}{\sqrt{1 + 10\omega_b^{3/4}}}~\mathrm{Mpc}/h.
 \end{aligned}
\end{equation}
 Note that this set of fitting functions applies strictly only to $\Lambda$CDM cosmologies.

\bibliographystyle{utcaps}
\bibliography{nuSpoon}

\providecommand{\href}[2]{#2}\begingroup\raggedright\begin{thebibliography}{10}

\bibitem{Tanabashi:2018oca}
{\bfseries Particle Data Group} Collaboration, M.~Tanabashi {\em et al.},
  ``{Review of Particle Physics},''
  \href{http://dx.doi.org/10.1103/PhysRevD.98.030001}{{\em Phys. Rev. D}
  {\bfseries 98} (2018) no.~3, 030001}.

\bibitem{deSalas:2017kay}
P.~de~Salas, D.~Forero, C.~Ternes, M.~T{\'o}rtola, and J.~Valle, ``{Status of
  neutrino oscillations 2018: 3$\sigma$ hint for normal mass ordering and
  improved CP sensitivity},''
  \href{http://dx.doi.org/10.1016/j.physletb.2018.06.019}{{\em Phys. Lett. B}
  {\bfseries 782} (2018)  633--640},
  \href{http://arxiv.org/abs/1708.01186}{{\ttfamily arXiv:1708.01186
  [hep-ph]}}.

\bibitem{Esteban:2018azc}
I.~Esteban, M.~Gonzalez-Garcia, A.~Hernandez-Cabezudo, M.~Maltoni, and
  T.~Schwetz, ``{Global analysis of three-flavour neutrino oscillations:
  synergies and tensions in the determination of $\theta_{23}$, $\delta_{CP}$,
  and the mass ordering},''
  \href{http://dx.doi.org/10.1007/JHEP01(2019)106}{{\em JHEP} {\bfseries 01}
  (2019)  106}, \href{http://arxiv.org/abs/1811.05487}{{\ttfamily
  arXiv:1811.05487 [hep-ph]}}.

\bibitem{Kraus:2004zw}
C.~Kraus {\em et al.}, ``{Final results from phase II of the Mainz neutrino
  mass search in tritium beta decay},''
  \href{http://dx.doi.org/10.1140/epjc/s2005-02139-7}{{\em Eur. Phys. J. C}
  {\bfseries 40} (2005)  447--468},
  \href{http://arxiv.org/abs/hep-ex/0412056}{{\ttfamily arXiv:hep-ex/0412056}}.

\bibitem{Aseev:2011dq}
{\bfseries Troitsk} Collaboration, V.~Aseev {\em et al.}, ``{An upper limit on
  electron antineutrino mass from Troitsk experiment},''
  \href{http://dx.doi.org/10.1103/PhysRevD.84.112003}{{\em Phys. Rev. D}
  {\bfseries 84} (2011)  112003},
  \href{http://arxiv.org/abs/1108.5034}{{\ttfamily arXiv:1108.5034 [hep-ex]}}.

\bibitem{Aker:2019uuj}
{\bfseries KATRIN} Collaboration, M.~Aker {\em et al.}, ``{Improved Upper Limit
  on the Neutrino Mass from a Direct Kinematic Method by KATRIN},''
  \href{http://dx.doi.org/10.1103/PhysRevLett.123.221802}{{\em Phys. Rev.
  Lett.} {\bfseries 123} (2019) no.~22, 221802},
  \href{http://arxiv.org/abs/1909.06048}{{\ttfamily arXiv:1909.06048
  [hep-ex]}}.

\bibitem{Lesgourgues:2012uu}
J.~Lesgourgues and S.~Pastor, ``{Neutrino mass from Cosmology},''
  \href{http://dx.doi.org/10.1155/2012/608515}{{\em Adv.\ High Energy Phys.}
  {\bfseries 2012} (2012)  608515},
  \href{http://arxiv.org/abs/1212.6154}{{\ttfamily arXiv:1212.6154 [hep-ph]}}.

\bibitem{Wong:2011ip}
Y.~Y. Wong, ``{Neutrino mass in cosmology: status and prospects},''
  \href{http://dx.doi.org/10.1146/annurev-nucl-102010-130252}{{\em Ann. Rev.
  Nucl. Part. Sci.} {\bfseries 61} (2011)  69--98},
  \href{http://arxiv.org/abs/1111.1436}{{\ttfamily arXiv:1111.1436
  [astro-ph.CO]}}.

\bibitem{Hannestad:2006zg}
S.~Hannestad, ``{Primordial neutrinos},''
  \href{http://dx.doi.org/10.1146/annurev.nucl.56.080805.140548}{{\em Ann. Rev.
  Nucl. Part. Sci.} {\bfseries 56} (2006)  137--161},
  \href{http://arxiv.org/abs/hep-ph/0602058}{{\ttfamily arXiv:hep-ph/0602058}}.

\bibitem{Lewis:1999bs}
A.~Lewis, A.~Challinor, and A.~Lasenby, ``{Efficient computation of CMB
  anisotropies in closed FRW models},''
  \href{http://dx.doi.org/10.1086/309179}{{\em Astrophys.\ J.} {\bfseries 538}
  (2000)  473--476},
\href{http://arxiv.org/abs/astro-ph/9911177}{{\ttfamily arXiv:astro-ph/9911177
  [astro-ph]}}.
%%CITATION = ASTRO-PH/9911177;%%.

\bibitem{Howlett:2012mh}
C.~Howlett, A.~Lewis, A.~Hall, and A.~Challinor, ``{CMB power spectrum
  parameter degeneracies in the era of precision cosmology},''
  \href{http://dx.doi.org/10.1088/1475-7516/2012/04/027}{{\em JCAP} {\bfseries
  1204} (2012)  027},
\href{http://arxiv.org/abs/1201.3654}{{\ttfamily arXiv:1201.3654
  [astro-ph.CO]}}.
%%CITATION = ARXIV:1201.3654;%%.

\bibitem{Lesgourgues:2011re}
J.~Lesgourgues, ``{The Cosmic Linear Anisotropy Solving System (CLASS) I:
  Overview},'' \href{http://arxiv.org/abs/1104.2932}{{\ttfamily arXiv:1104.2932
  [astro-ph.IM]}}.

\bibitem{Upadhye:2017hdl}
A.~Upadhye, ``{Neutrino mass and dark energy constraints from redshift-space
  distortions},'' \href{http://dx.doi.org/10.1088/1475-7516/2019/05/041}{{\em
  JCAP} {\bfseries 05} (2019)  041},
  \href{http://arxiv.org/abs/1707.09354}{{\ttfamily arXiv:1707.09354
  [astro-ph.CO]}}.

\bibitem{Aghanim:2018eyx}
{\bfseries Planck} Collaboration, N.~Aghanim {\em et al.}, ``{Planck 2018
  results. VI. Cosmological parameters},''
  \href{http://arxiv.org/abs/1807.06209}{{\ttfamily arXiv:1807.06209
  [astro-ph.CO]}}.

\bibitem{RoyChoudhury:2019hls}
S.~Roy~Choudhury and S.~Hannestad, ``{Updated results on neutrino mass and mass
  hierarchy from cosmology with Planck 2018 likelihoods},''
  \href{http://arxiv.org/abs/1907.12598}{{\ttfamily arXiv:1907.12598
  [astro-ph.CO]}}.

\bibitem{Hamann:2012fe}
J.~Hamann, S.~Hannestad, and Y.~Y. Wong, ``{Measuring neutrino masses with a
  future galaxy survey},''
  \href{http://dx.doi.org/10.1088/1475-7516/2012/11/052}{{\em JCAP} {\bfseries
  11} (2012)  052}, \href{http://arxiv.org/abs/1209.1043}{{\ttfamily
  arXiv:1209.1043 [astro-ph.CO]}}.

\bibitem{Amendola:2016saw}
L.~Amendola {\em et al.}, ``{Cosmology and fundamental physics with the Euclid
  satellite},'' \href{http://dx.doi.org/10.1007/s41114-017-0010-3}{{\em Living
  Rev. Rel.} {\bfseries 21} (2018) no.~1, 2},
  \href{http://arxiv.org/abs/1606.00180}{{\ttfamily arXiv:1606.00180
  [astro-ph.CO]}}.

\bibitem{Brandbyge:2008rv}
J.~Brandbyge, S.~Hannestad, T.~Haugb{\o}lle, and B.~Thomsen, ``{The Effect of
  Thermal Neutrino Motion on the Non-linear Cosmological Matter Power
  Spectrum},'' \href{http://dx.doi.org/10.1088/1475-7516/2008/08/020}{{\em
  JCAP} {\bfseries 08} (2008)  020},
  \href{http://arxiv.org/abs/0802.3700}{{\ttfamily arXiv:0802.3700
  [astro-ph]}}.

\bibitem{Brandbyge:2008js}
J.~Brandbyge and S.~Hannestad, ``{Grid Based Linear Neutrino Perturbations in
  Cosmological N-body Simulations},''
  \href{http://dx.doi.org/10.1088/1475-7516/2009/05/002}{{\em JCAP} {\bfseries
  05} (2009)  002}, \href{http://arxiv.org/abs/0812.3149}{{\ttfamily
  arXiv:0812.3149 [astro-ph]}}.

\bibitem{Brandbyge:2009ce}
J.~Brandbyge and S.~Hannestad, ``{Resolving Cosmic Neutrino Structure: A Hybrid
  Neutrino N-body Scheme},''
  \href{http://dx.doi.org/10.1088/1475-7516/2010/01/021}{{\em JCAP} {\bfseries
  01} (2010)  021}, \href{http://arxiv.org/abs/0908.1969}{{\ttfamily
  arXiv:0908.1969 [astro-ph.CO]}}.

\bibitem{Viel:2010bn}
M.~Viel, M.~G. Haehnelt, and V.~Springel, ``{The effect of neutrinos on the
  matter distribution as probed by the Intergalactic Medium},''
  \href{http://dx.doi.org/10.1088/1475-7516/2010/06/015}{{\em JCAP} {\bfseries
  06} (2010)  015}, \href{http://arxiv.org/abs/1003.2422}{{\ttfamily
  arXiv:1003.2422 [astro-ph.CO]}}.

\bibitem{Bird:2011rb}
S.~Bird, M.~Viel, and M.~G. Haehnelt, ``{Massive Neutrinos and the Non-linear
  Matter Power Spectrum},''
  \href{http://dx.doi.org/10.1111/j.1365-2966.2011.20222.x}{{\em Mon. Not. Roy.
  Astron. Soc.} {\bfseries 420} (2012)  2551--2561},
  \href{http://arxiv.org/abs/1109.4416}{{\ttfamily arXiv:1109.4416
  [astro-ph.CO]}}.

\bibitem{AliHaimoud:2012vj}
Y.~Ali-Ha{\"\i}moud and S.~Bird, ``{An efficient implementation of massive
  neutrinos in non-linear structure formation simulations},''
  \href{http://dx.doi.org/10.1093/mnras/sts286}{{\em Mon. Not. Roy. Astron.
  Soc.} {\bfseries 428} (2012)  3375--3389},
  \href{http://arxiv.org/abs/1209.0461}{{\ttfamily arXiv:1209.0461
  [astro-ph.CO]}}.

\bibitem{Castorina:2015bma}
E.~Castorina, C.~Carbone, J.~Bel, E.~Sefusatti, and K.~Dolag, ``{DEMNUni: The
  clustering of large-scale structures in the presence of massive neutrinos},''
  \href{http://dx.doi.org/10.1088/1475-7516/2015/07/043}{{\em JCAP} {\bfseries
  07} (2015)  043}, \href{http://arxiv.org/abs/1505.07148}{{\ttfamily
  arXiv:1505.07148 [astro-ph.CO]}}.

\bibitem{Banerjee:2016zaa}
A.~Banerjee and N.~Dalal, ``{Simulating nonlinear cosmological structure
  formation with massive neutrinos},''
  \href{http://dx.doi.org/10.1088/1475-7516/2016/11/015}{{\em JCAP} {\bfseries
  11} (2016)  015}, \href{http://arxiv.org/abs/1606.06167}{{\ttfamily
  arXiv:1606.06167 [astro-ph.CO]}}.

\bibitem{Liu:2017now}
J.~Liu, S.~Bird, J.~M.~Z. Matilla, J.~C. Hill, Z.~Haiman, M.~S. Madhavacheril,
  A.~Petri, and D.~N. Spergel, ``{MassiveNuS: Cosmological Massive Neutrino
  Simulations},'' \href{http://dx.doi.org/10.1088/1475-7516/2018/03/049}{{\em
  JCAP} {\bfseries 03} (2018)  049},
  \href{http://arxiv.org/abs/1711.10524}{{\ttfamily arXiv:1711.10524
  [astro-ph.CO]}}.

\bibitem{Banerjee:2018bxy}
A.~Banerjee, D.~Powell, T.~Abel, and F.~Villaescusa-Navarro, ``{Reducing Noise
  in Cosmological N-body Simulations with Neutrinos},''
  \href{http://dx.doi.org/10.1088/1475-7516/2018/09/028}{{\em JCAP} {\bfseries
  09} (2018)  028}, \href{http://arxiv.org/abs/1801.03906}{{\ttfamily
  arXiv:1801.03906 [astro-ph.CO]}}.

\bibitem{Partmann:2020qzb}
C.~Partmann, C.~Fidler, C.~Rampf, and O.~Hahn, ``{Fast simulations of cosmic
  large-scale structure with massive neutrinos},''
  \href{http://arxiv.org/abs/2003.07387}{{\ttfamily arXiv:2003.07387
  [astro-ph.CO]}}.

\bibitem{Wong:2008ws}
Y.~Y. Wong, ``{Higher order corrections to the large scale matter power
  spectrum in the presence of massive neutrinos},''
  \href{http://dx.doi.org/10.1088/1475-7516/2008/10/035}{{\em JCAP} {\bfseries
  10} (2008)  035}, \href{http://arxiv.org/abs/0809.0693}{{\ttfamily
  arXiv:0809.0693 [astro-ph]}}.

\bibitem{Fuhrer:2014zka}
F.~F{\"u}hrer and Y.~Y.~Y. Wong, ``{Higher-order massive neutrino perturbations
  in large-scale structure},''
  \href{http://dx.doi.org/10.1088/1475-7516/2015/03/046}{{\em JCAP} {\bfseries
  03} (2015)  046}, \href{http://arxiv.org/abs/1412.2764}{{\ttfamily
  arXiv:1412.2764 [astro-ph.CO]}}.

\bibitem{Lesgourgues:2009am}
J.~Lesgourgues, S.~Matarrese, M.~Pietroni, and A.~Riotto, ``{Non-linear Power
  Spectrum including Massive Neutrinos: the Time-RG Flow Approach},''
  \href{http://dx.doi.org/10.1088/1475-7516/2009/06/017}{{\em JCAP} {\bfseries
  06} (2009)  017}, \href{http://arxiv.org/abs/0901.4550}{{\ttfamily
  arXiv:0901.4550 [astro-ph.CO]}}.

\bibitem{Blas:2014hya}
D.~Blas, M.~Garny, T.~Konstandin, and J.~Lesgourgues, ``{Structure formation
  with massive neutrinos: going beyond linear theory},''
  \href{http://dx.doi.org/10.1088/1475-7516/2014/11/039}{{\em JCAP} {\bfseries
  11} (2014)  039}, \href{http://arxiv.org/abs/1408.2995}{{\ttfamily
  arXiv:1408.2995 [astro-ph.CO]}}.

\bibitem{Upadhye:2015lia}
A.~Upadhye, J.~Kwan, A.~Pope, K.~Heitmann, S.~Habib, H.~Finkel, and
  N.~Frontiere, ``{Redshift-space distortions in massive neutrino and evolving
  dark energy cosmologies},''
  \href{http://dx.doi.org/10.1103/PhysRevD.93.063515}{{\em Phys.\ Rev.\ D}
  {\bfseries 93} (2016) no.~6, 063515},
  \href{http://arxiv.org/abs/1506.07526}{{\ttfamily arXiv:1506.07526
  [astro-ph.CO]}}.

\bibitem{Pedersen:2019ieb}
C.~Pedersen, A.~Font-Ribera, T.~D. Kitching, P.~McDonald, S.~Bird, A.~z.
  Slosar, K.~K. Rogers, and A.~Pontzen, ``{Massive neutrinos and degeneracies
  in Lyman-alpha forest simulations},''
  \href{http://dx.doi.org/10.1088/1475-7516/2020/04/025}{{\em JCAP} {\bfseries
  04} (2020)  025}, \href{http://arxiv.org/abs/1911.09596}{{\ttfamily
  arXiv:1911.09596 [astro-ph.CO]}}.

\bibitem{Hannestad:2019piu}
S.~Hannestad and Y.~Y. Wong, ``{Fitting functions on the cheap: the relative
  nonlinear matter power spectrum},''
  \href{http://dx.doi.org/10.1088/1475-7516/2020/03/028}{{\em JCAP} {\bfseries
  03} (2020)  028}, \href{http://arxiv.org/abs/1907.01125}{{\ttfamily
  arXiv:1907.01125 [astro-ph.CO]}}.

\bibitem{Cooray:2002dia}
A.~Cooray and R.~K. Sheth, ``{Halo Models of Large Scale Structure},''
  \href{http://dx.doi.org/10.1016/S0370-1573(02)00276-4}{{\em Phys.\ Rept.}
  {\bfseries 372} (2002)  1--129},
  \href{http://arxiv.org/abs/astro-ph/0206508}{{\ttfamily
  arXiv:astro-ph/0206508}}.

\bibitem{Massara:2014kba}
E.~Massara, F.~Villaescusa-Navarro, and M.~Viel, ``{The halo model in a massive
  neutrino cosmology},''
  \href{http://dx.doi.org/10.1088/1475-7516/2014/12/053}{{\em JCAP} {\bfseries
  12} (2014)  053}, \href{http://arxiv.org/abs/1410.6813}{{\ttfamily
  arXiv:1410.6813 [astro-ph.CO]}}.

\bibitem{LoVerde:2014rxa}
M.~LoVerde, ``{Spherical collapse in $\nu \Lambda$CDM},''
  \href{http://dx.doi.org/10.1103/PhysRevD.90.083518}{{\em Phys. Rev. D}
  {\bfseries 90} (2014) no.~8, 083518},
  \href{http://arxiv.org/abs/1405.4858}{{\ttfamily arXiv:1405.4858
  [astro-ph.CO]}}.

\bibitem{Basse:2010qp}
T.~Basse, O.~E. Bjaelde, and Y.~Y. Wong, ``{Spherical collapse of dark energy
  with an arbitrary sound speed},''
  \href{http://dx.doi.org/10.1088/1475-7516/2011/10/038}{{\em JCAP} {\bfseries
  10} (2011)  038}, \href{http://arxiv.org/abs/1009.0010}{{\ttfamily
  arXiv:1009.0010 [astro-ph.CO]}}.

\bibitem{Navarro:1996gj}
J.~F. Navarro, C.~S. Frenk, and S.~D. White, ``{A Universal density profile
  from hierarchical clustering},'' \href{http://dx.doi.org/10.1086/304888}{{\em
  Astrophys.\ J.} {\bfseries 490} (1997)  493--508},
  \href{http://arxiv.org/abs/astro-ph/9611107}{{\ttfamily
  arXiv:astro-ph/9611107}}.

\bibitem{Bullock:1999he}
J.~S. Bullock, T.~S. Kolatt, Y.~Sigad, R.~S. Somerville, A.~V. Kravtsov, A.~A.
  Klypin, J.~R. Primack, and A.~Dekel, ``{Profiles of dark haloes. Evolution,
  scatter, and environment},''
  \href{http://dx.doi.org/10.1046/j.1365-8711.2001.04068.x}{{\em Mon. Not. Roy.
  Astron. Soc.} {\bfseries 321} (2001)  559--575},
\href{http://arxiv.org/abs/astro-ph/9908159}{{\ttfamily arXiv:astro-ph/9908159
  [astro-ph]}}.
%%CITATION = ASTRO-PH/9908159;%%.

\bibitem{Kwan:2012nd}
J.~Kwan, S.~Bhattacharya, K.~Heitmann, and S.~Habib, ``{Cosmic Emulation: The
  Concentration-Mass Relation for wCDM Universes},''
  \href{http://dx.doi.org/10.1088/0004-637X/768/2/123}{{\em Astrophys.\ J.}
  {\bfseries 768} (2013)  123},
  \href{http://arxiv.org/abs/1210.1576}{{\ttfamily arXiv:1210.1576
  [astro-ph.CO]}}.

\bibitem{Brandbyge:2010ge}
J.~Brandbyge, S.~Hannestad, T.~Haugb{\o}lle, and Y.~Y. Wong, ``{Neutrinos in
  Non-linear Structure Formation - The Effect on Halo Properties},''
  \href{http://dx.doi.org/10.1088/1475-7516/2010/09/014}{{\em JCAP} {\bfseries
  09} (2010)  014}, \href{http://arxiv.org/abs/1004.4105}{{\ttfamily
  arXiv:1004.4105 [astro-ph.CO]}}.

\bibitem{Bhattacharya:2010wy}
S.~Bhattacharya, K.~Heitmann, M.~White, Z.~Lukic, C.~Wagner, and S.~Habib,
  ``{Mass Function Predictions Beyond LCDM},''
  \href{http://dx.doi.org/10.1088/0004-637X/732/2/122}{{\em Astrophys.\ J.}
  {\bfseries 732} (2011)  122},
  \href{http://arxiv.org/abs/1005.2239}{{\ttfamily arXiv:1005.2239
  [astro-ph.CO]}}.

\bibitem{Biswas:2019uhy}
R.~Biswas, K.~Heitmann, S.~Habib, A.~Upadhye, A.~Pope, and N.~Frontiere,
  ``{Effects of Massive Neutrinos and Dynamical Dark Energy on the Cluster Mass
  Function},'' \href{http://arxiv.org/abs/1901.10690}{{\ttfamily
  arXiv:1901.10690 [astro-ph.CO]}}.

\bibitem{Press:1973iz}
W.~H. Press and P.~Schechter, ``{Formation of galaxies and clusters of galaxies
  by selfsimilar gravitational condensation},''
  \href{http://dx.doi.org/10.1086/152650}{{\em Astrophys.\ J.} {\bfseries 187}
  (1974)  425--438}.

\bibitem{Sheth:1999mn}
R.~K. Sheth and G.~Tormen, ``{Large scale bias and the peak background
  split},'' \href{http://dx.doi.org/10.1046/j.1365-8711.1999.02692.x}{{\em
  Mon.\ Not.\ Roy.\ Astron.\ Soc.} {\bfseries 308} (1999)  119},
  \href{http://arxiv.org/abs/astro-ph/9901122}{{\ttfamily
  arXiv:astro-ph/9901122}}.

\bibitem{Heitmann:2013bra}
K.~Heitmann, E.~Lawrence, J.~Kwan, S.~Habib, and D.~Higdon, ``{The Coyote
  Universe Extended: Precision Emulation of the Matter Power Spectrum},''
  \href{http://dx.doi.org/10.1088/0004-637X/780/1/111}{{\em Astrophys. J.}
  {\bfseries 780} (2014)  111},
\href{http://arxiv.org/abs/1304.7849}{{\ttfamily arXiv:1304.7849
  [astro-ph.CO]}}.
%%CITATION = ARXIV:1304.7849;%%.

\bibitem{Eisenstein:1997ik}
D.~J. Eisenstein and W.~Hu, ``{Baryonic features in the matter transfer
  function},'' \href{http://dx.doi.org/10.1086/305424}{{\em Astrophys. J.}
  {\bfseries 496} (1998)  605},
  \href{http://arxiv.org/abs/astro-ph/9709112}{{\ttfamily
  arXiv:astro-ph/9709112}}.

\bibitem{Potter:2016ttn}
D.~Potter, J.~Stadel, and R.~Teyssier, ``{PKDGRAV3: Beyond Trillion Particle
  Cosmological Simulations for the Next Era of Galaxy Surveys},''
  \href{http://arxiv.org/abs/1609.08621}{{\ttfamily arXiv:1609.08621
  [astro-ph.IM]}}.

\bibitem{Dakin:2017idt}
J.~Dakin, J.~Brandbyge, S.~Hannestad, T.~Haugb{\o}lle, and T.~Tram,
  ``{$\nu$CO$N$CEPT: Cosmological neutrino simulations from the non-linear
  Boltzmann hierarchy},''
  \href{http://dx.doi.org/10.1088/1475-7516/2019/02/052}{{\em JCAP} {\bfseries
  02} (2019)  052}, \href{http://arxiv.org/abs/1712.03944}{{\ttfamily
  arXiv:1712.03944 [astro-ph.CO]}}.

\bibitem{Pietroni:2008jx}
M.~Pietroni, ``{Flowing with Time: a New Approach to Nonlinear Cosmological
  Perturbations},'' \href{http://dx.doi.org/10.1088/1475-7516/2008/10/036}{{\em
  JCAP} {\bfseries 10} (2008)  036},
  \href{http://arxiv.org/abs/0806.0971}{{\ttfamily arXiv:0806.0971
  [astro-ph]}}.

\bibitem{Upadhye:2013ndm}
A.~Upadhye, R.~Biswas, A.~Pope, K.~Heitmann, S.~Habib, H.~Finkel, and
  N.~Frontiere, ``{Large-Scale Structure Formation with Massive Neutrinos and
  Dynamical Dark Energy},''
  \href{http://dx.doi.org/10.1103/PhysRevD.89.103515}{{\em Phys. Rev. D}
  {\bfseries 89} (2014) no.~10, 103515},
  \href{http://arxiv.org/abs/1309.5872}{{\ttfamily arXiv:1309.5872
  [astro-ph.CO]}}.

\bibitem{McDonald:2006mx}
P.~McDonald, ``{Clustering of dark matter tracers: Renormalizing the bias
  parameters},'' \href{http://dx.doi.org/10.1103/PhysRevD.74.129901}{{\em Phys.
  Rev. D} {\bfseries 74} (2006)  103512},
  \href{http://arxiv.org/abs/astro-ph/0609413}{{\ttfamily
  arXiv:astro-ph/0609413}}. [Erratum: Phys.Rev.D 74, 129901 (2006)].

\bibitem{Nishizawa:2012db}
A.~J. Nishizawa, M.~Takada, and T.~Nishimichi, ``{Perturbation theory for
  nonlinear halo power spectrum: the renormalized bias and halo bias},''
  \href{http://dx.doi.org/10.1093/mnras/stt716}{{\em Mon. Not. Roy. Astron.
  Soc.} {\bfseries 433} (2013)  209},
  \href{http://arxiv.org/abs/1212.4025}{{\ttfamily arXiv:1212.4025
  [astro-ph.CO]}}.

\bibitem{Desjacques:2016bnm}
V.~Desjacques, D.~Jeong, and F.~Schmidt, ``{Large-Scale Galaxy Bias},''
  \href{http://dx.doi.org/10.1016/j.physrep.2017.12.002}{{\em Phys. Rept.}
  {\bfseries 733} (2018)  1--193},
  \href{http://arxiv.org/abs/1611.09787}{{\ttfamily arXiv:1611.09787
  [astro-ph.CO]}}.

\end{thebibliography}\endgroup

\end{document}